\documentclass[twocolumn,epjc3]{svjour3}
\smartqed  % flush right qed marks, e.g. at end of proof
\usepackage[english]{babel}
\usepackage{epsfig}% Include figure files
\usepackage{graphicx}
\usepackage{amssymb}
\usepackage{amsfonts}
\usepackage{amsmath}
\usepackage{longtable}
\usepackage{rotating}
\usepackage{lscape}
\usepackage{multirow}
\usepackage{float}
\usepackage{bm}

\journalname{Eur. Phys. J. A}

\begin{document}

\title{Properties of dibaryons in nuclear medium}
\author{M. Kakenov\thanksref{e1,addr1,addr11}
\and
V.I. Kukulin\thanksref{e2,addr2}
\and
V.N. Pomerantsev\thanksref{e3,addr2}
\and
O. Bayakhmetov\thanksref{e4,addr4}
}
\thankstext{e1}{e-mail:kakenov@jinr.ru}
\thankstext{e2}{e-mail:kukulin@nucl-th.sinp.msu.ru}
\thankstext{e3}{e-mail:pomeran@nucl-th.sinp.msu.ru}
\thankstext{e4}{e-mail:bayakhmetov.o.s.92@gmail.com}

\institute{Laboratory of Information Technologies, Joint Institute for Nuclear Research, 6 Joliot-Curie, Dubna, Moscow region, 141980, Russia\label{addr1}
\and
Institute of Nuclear Physics, 1 Ibragimova street, Almaty, 050032, Kazakhstan\label{addr11}
\and
Skobeltsyn Institute of Nuclear Physics, Lomonosov Moscow State
University, 1(2) Leninskie Gory, Moscow, 119991, Russia\label{addr2}
\and
L.N. Gumilyov Eurasian National University, 2 Satpaev Str, Nur-Sultan, 010008 Kazakhstan\label{addr4}
}
\date{Received: \today / Accepted: date}

\maketitle

\begin{abstract} 
Properties of six-quark dibaryons in nuclear medium are
considered by example of $A=6$ nuclei within the three-cluster $\alpha+2N$ model.
Dibaryon production in nuclei leads to the appearance of a three-body force
between the dibaryon and nuclear core. This non-conventional scalar force is shown to provide an additional attractive contribution to the three-body binding energy.
This three-body contribution improves noticeably agreement between
theoretical results and experimental data for the majority of observables. The
most serious difference between the traditional $NN$-force models and the
dibaryon-induced model is found for the nucleon momentum distribution, the
latter model providing a strong enrichment of the high-momentum components both for $^6$Li and $^6$He cases.

%\PACS{03.65.Nk,21.45.-v,25.45.De}
\end{abstract}

\section{Introduction. What is dibaryon concept for nuclear force?}
The history of the nuclear force, i.e., the force which holds the nucleons in a
nucleus together, is very long, contradictory and dramatic. It started almost
immediately after the discovery of the neutron by J. Chadwick in 1932 and even now
it can in no way be considered as completed. The dominant paradigm suggested by
H. Yukawa in 1935 \cite{Yukawa} is based on a general idea of pion exchange
between nucleons both in free space and inside nuclei. This basic concept
being generalized later to include other meson exchanges \cite{Breit,Signell,Bonn_NN}
was able to explain many fundamental properties of nuclear interactions like their
short-range character, presence of tensor and spin-orbit forces, properties of
the deuteron and light nuclei, etc. But simultaneously this model met very numerous
contradictions and inconsistencies when interpreting the individual nuclear
phenomena \cite{5,6,7,8,9,10}.
Therefore, to put the nuclear theory on the more consistent and solid ground, the so-called realistic one-boson exchange (OBE)
$NN$ potentials used conventionally for numerous nuclear phenomena calculations have been replaced in recent years by the potentials derived
from the Effective Field Theory (EFT), or the Chiral Perturbation Theory (ChPT)
\cite{Weinberg,Machl,Eppelbaum,Ordones,Hammer}. Lately, the EFT approach has attracted
a strong interest of the researchers in the field and it gave birth to the hopes
to explain systematically many puzzles and paradoxes existing in nuclear physics at all.

The most convenient and appropriate field to apply the refined and powerful EFT-based approaches are the few-nucleon (and, in general, few-body) systems which could be treated accurately, without further approximations using the well-known Faddeev equation formalism.

It is well known that there are numerous and long-standing puzzles and paradoxes in this field, where approaches based on phenomenological and semi-pheno\-meno\-logi\-cal realistic $NN$ potentials have failed in explaining the experimental data. Among such puzzles one can enumerate the following ones:

-- $A_y$ puzzle in the vector analyzing power for low-energy scattering of polarized nucleons on deuterons $\vec{N}+d$ and also puzzles in the deuteron tensor analyzing powers $T_{20}$, $T_{21}$ and $T_{22}$ for scattering of nucleons on polarized deuterons $N+\vec{d}$~\cite{5,6};

-- the Sagara puzzle in the $N+d$ elastic scattering cross section at low energies and large angles~\cite{8};

-- the space-star three-nucleon breakup in $n+d$ and $p+d$ collisions;

-- the differential cross section and vector and tensor analyzing powers for $n+d$ and $p+d$ backward scattering at intermediate energies 200--400 MeV~\cite{9};

-- Coulomb effects in the three-body breakup in $N+d$ collisions, etc.

None of these puzzles has found a satisfactory explanation when passing from the traditional meson-exchange realistic $NN$
and $3N$ potentials to the modern EFT-based approaches\footnote{%
According to the recent study~\cite{Margaryan} of the low-energy analyzing powers, the values of $A_y$, $iT_{11}$ and $T_{20}$ can be explained only by varying the contact $P$-wave terms within the N$^3$LO and N$^4$LO approximations.}. So, the origin for
the failure in explaining the above puzzling phenomena appears to be the absence of some key ingredients both in the traditional meson-exchange $NN$ potentials and in the EFT-based interaction.

There is another class of phenomena where the traditional meson-exchange and the modern EFT
approaches face serious conceptual problems. These are processes accompanied by high momentum
transfers or occurring at short internucleon distances, like:

-- the so-called cumulative particle production, i.e., production in the
kinematical region forbidden for a single-particle process \cite{Leksin,16,17,18};

-- the sub-threshold particle production \cite{19,20,21}, etc.

Here one should mention also a very interesting feature discovered only recently: a direct interrelation between the
short-range $NN$ correlations in nuclei and the single-quark momentum distributions found in such phenomena as Deep Inelastic Scattering (DIS), EMC effect, etc.~\cite{Hen,23,24}.

A possible explanation for these phenomena can be related to the production
of some intermediate strongly correlated two- and few-nucleon clusters in nuclei
(the so-called di- and multibaryons) which behave like tightly bound quasi-particles
and can absorb a very large momentum from the high-energy projectile. In turn,
such a two- or few-nucleon clustering is tightly related to the well-known
short-range correlations (SRC) of nucleons in nuclei which have been studied
extensively in the numerous experiments done in JLab for the last two decades (see, e.g.,
\cite{Hen,23,24}). These SRC are apparently a manifestation of the quark
structure of the nucleon and some features of the interquark interactions, which are
described in the framework of the Quantum Chromodynamics (QCD).

It is commonly accepted that the QCD gives the general basis for nuclear physics. However,
the QCD operates with such degrees of freedom (d.o.f.) --- quarks, gluons, strings ---
that are completely different from those with which traditional nuclear physics
deals. Hence, when describing the nuclear phenomena with a high-momentum transfer, one
needs a bridge between the fundamental QCD and the traditional nuclear physical
d.o.f. The conventional meson-exchange and EFT approaches do not provide
such a bridge.

Thus, to move further one needs a sort of hybrid model combining both
quark-gluon and meson-exchange aspects of the $NN$ interaction. The specific model
for the nuclear force which unifies both quark-gluon and meson-exchange features --- the dibaryon (or dibaryon-induced) model --- was developed by Kukulin et al. in the last two decades \cite{PIYAF,YaF2001,JPhys2001,KuInt,Yaf2019,PLB,EPJA} (in the early studies~\cite{PIYAF,YaF2001,JPhys2001,KuInt} it was called ``the
dressed-bag model''). The model allowed for a very good description of both elastic
and inelastic $NN$ scattering phase shifts from zero energy up to $T_{\rm lab}=600$--$800$ MeV (or even up to 1~GeV in some partial waves)~\cite{Yaf2019,PLB,EPJA} and the
deuteron properties~\cite{KuInt} using only a few basic parameters.
Moreover, when considering the three-nucleon system within the framework of the dibaryon model, a specific three-particle force inevitably arises due to the interaction of an intermediate dibaryon formed from a pair of nucleons with the third nucleon \cite{sys3n,YAF3N}. Our first
calculations of the ground states of the three-nucleon nuclei $^3$H and $^3$He have
shown that the dibaryon model for $2N$ and $3N$ forces gives a good description
of the basic properties of these nuclei, including the precise value for the
Coulomb displacement energy $\Delta E_{\rm C}$ \cite{sys3n}.

Nevertheless, since our approach is highly non-canonical, it should be tested
carefully in other nuclear physics predictions. So, in the present paper we have chosen
for such a test the properties of the $A=6$ nuclei within the three-cluster
$\alpha+2N$ model. These nuclei ($^6$He and $^6$Li) are very well suited for
our purposes because their structure can be described quite well within the model of two
interacting nucleons in the field of the inert core \cite{voronchev82,NP84,NP86,NP90,NP93,NP95}.

So, here we suggest to compare three groups of predictions:

--  properties of the $A=6$ nuclei with the conventional $NN$ potentials;

--  properties of the $A=6$ nuclei with the dibaryon model potential for the $NN$ and
respective three-body interactions;

-- properties of the isolated $NN$ system, i.e., when the $\alpha$-particle core is removed.

This comparison allows to shed light not only on the properties of dibaryons in
nuclear systems but also on some key features of specific nuclear phenomena such
as pairing and the nature of Cooper pairs in nuclei. We plan to discuss these
interesting problems in a separate paper.

A few words should be said about the impact of some fine effects, such as relativistic effects and possible excitation of the alpha cluster core. The main focus of the present study is to investigate the behaviour and properties of dibaryons in a nuclear field by comparing the properties of the same nuclei within two alternative $NN$-force models, namely the traditional one and that based on the dibaryon concept. So, one can hope that the differences between the results obtained for these two types of force models are hardly sensitive to the above finer corrections. Moreover, the accurate $3N$ calculations of the Bochum--Cracow group \cite{Rel} and also our estimations showed that the relativistic effects for such {\em differences} are almost negligible.

The structure of the paper is as follows. In Sec.~2, we discuss briefly the three-cluster $\alpha+2N$ model for the $A=6$ nuclei and present the variational formalism for the calculations of these nuclei.
In Sec.~3 we describe the dibaryon model for the $NN$ interaction and its employment for the basic $NN$ channels in the $\alpha+2N$ system. Here we also introduce the interactions between the intermediate dibaryon and the third nucleon in the $3N$ system or the $\alpha$-core in the $\alpha+2N$ system which result in three-body forces for both systems.
In Sec.~4 we present the results of our detailed calculations for the ground states of the $A=6$ nuclei within the dibaryon-induced model for the $NN$ and $NN\alpha$ forces in comparison with the results for the traditional Reid Soft Core (RSC) model.
Here we discuss some observables for the $A=6$ nuclei with the focus on the momentum distributions at high momenta and discuss the relation of our results to some fundamental properties of nuclei at all, such as pairing of neutrons in the $^1S_0$ channel.
In Sec.~5 we study the dependence of the dibaryon admixture on nuclear density by changing the binding energy of the $NN$ pair in the nucleus due to changing the coupling constant for the three-body force.
In Sec. 6 we summarize the results and present some concluding remarks.

\section{Three-body $\alpha+2N$ cluster model for $A=6$ nuclei}
The three-body cluster model for the $A=6$ nuclei in its modern form was developed
first in a series of papers \cite{voronchev82,NP84,NP86,NP90,NP93,NP95}, see also Refs.~\cite{Danilin,40,41}\footnote{%
The previous version of the three-body model for the
$^{6}$Li nucleus~\cite{Lawson} was only schematic and could not lead to the quantitative and consistent predictions.}.
This model was able to explain in a quantitative manner many properties of the $A=6$ nuclei, though some minor disagreement with the data still remained, e.g., for the binding energies of $^{6}$He and $^{6}$Li (disagreement with the data was about 0.5--0.8 MeV) and for the rms charge radii (at the 5\% error level).
The physical justification for the three-cluster $\alpha+2N$ model for the $A=6$ nuclei ($^{6}$Li -- $^{6}$He -- $^{6}$Be) can be their low binding energy in the channel $\alpha+N+N$ or $\alpha+2N$ as compared to the binding energy of the $\alpha$-cluster.
Another important hint for the validity of the three-cluster model for the $A=6$ nuclei is the rather large interparticle distance in the ground and excited states of these nuclei, larger than the interparticle potential range (see the detailed discussion for these problems in Sec. 5). So, it is plausible to assume that when combining three particles $\alpha+N+N$ into an eigenstate of the above $A=6$ nuclei, the $\alpha$-cluster distortion is small and can be ignored.

Recently, within the similar approximations, an interesting EFT-based model for the $^6$He nucleus has been developed~\cite{Charl}. The approach employs basically the halo structure for $^6$He with three well-separated particles. Unfortunately, the authors of Ref.~\cite{Charl} studied only the binding energy and the gross structure of the $^6$He ground state within their EFT-Halo approach. Contrary to this, in the present work we study the basic properties of the $A=6$ nuclei, spin-orbit structure of the wave functions, their geometrical forms, nucleon momentum distributions, etc.

The three-body Hamiltonian for the $A=6$ nuclei includes the $\alpha N$ potential with the $0s$ forbidden
(by the Pauli principle) state \cite{NP95} and some realistic $NN$ potentials which are chosen here in
two alternative forms:\\
 -- in a form of a traditional $NN$ potential with the repulsive core (here we have chosen the RSC potential as in Ref.~\cite{NP95});\\
 -- in a form of the dibaryon-induced $NN$ potential \cite{KuInt}, which can describe the $NN$ scattering phase shifts from zero
energy up to $T_{\rm lab} \simeq 1$ GeV or even higher.

Keeping in mind that the RSC potential
can fit the empirical $NN$ phase shifts until $T_{\rm lab} = 300$ MeV only, while the dibaryon-induced $NN$ interaction can describe $NN$ scattering in a much broader energy range, one can assume that the off-shell properties of the dibaryon-induced potential should be much more adequate than those for the RSC model.

The main formal difference between the dibaryon-induced interaction and the
traditional $NN$-force model is its energy dependence which arises due to the exclusion of the internal
(dibaryon) channel. In the three-particle system $N+N+\alpha$, this energy
dependence also leads to the dependence of the pair $NN$ potential on the
momentum of the $\alpha$-particle relative to the center of mass of the $NN$
pair. Another feature of the dibaryon model is the presence of the three-particle
force due to the interaction of the dibaryon with the $\alpha$-particle. This
three-particle force in the effective Hamiltonian also depends on the energy of
the system.

\begin{figure}[h]
\begin{center}
\includegraphics[width=0.35\textwidth]{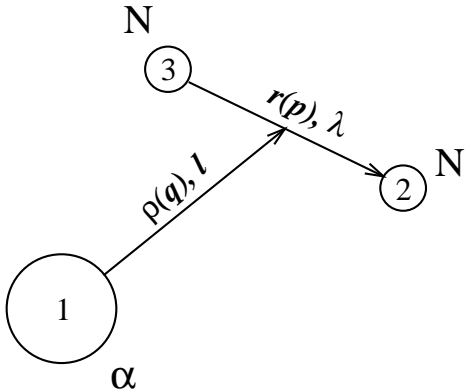}
\caption{Yacobi coordinates ${\bf r}$ and ${\bm \rho}$, their conjugated momenta ${\bm p}$ and ${\bf q}$, and respective angular momenta ${\bm \lambda}$ and ${\bm l}$ for the $\alpha+2N$ system. \label{Yacobi}}
\end{center}
\end{figure}

We use the following notation for coordinates and momenta (see
Fig. \ref{Yacobi}): ${\bf r}$ is the relative coordinate of
two nucleons, while ${\bm \rho}$ is the Jacobi coordinate of
the $\alpha$-particle relative to the center of mass of the nucleon pair; ${\bf p}$ and ${\bf q}$ are the momenta canonically conjugated to the coordinates ${\bf r}$ and ${\bm \rho}$, respectively. The
composite index $\gamma = \{\lambda,l,L,S\}$ represents the set of quantum
numbers for the basis functions: the orbital angular momenta $\lambda$ and $l$
correspond to the Jacobi coordinates $\bf r$  and $\bm\rho$, respectively,
${\bm L}={\bm l}+{\bm \lambda}$
is the total orbital angular momentum and $S$ is the spin of the system which in this case is equal to the total spin of two nucleons. The total angular momentum is ${\bf J} = {\bf L}+{\bm S}$.

For the case of the dibaryon-induced $NN$ potential the total effective three-body
Hamiltonian for the $\alpha+2N$ system takes the form:
\begin{equation} H^{\rm
tot}(E)= T+V_{N_2\alpha}+V_{N_3\alpha}+V_{NN}^{\rm eff}(E) + W_{\rm 3BF}(E)
\label{htot} \end{equation}
Here $T$ is the kinetic energy, $V_{N_{i}\alpha}$ is the potential of the interaction of the $i$-th nucleon with the $\alpha$-core, $V_{NN}^{\rm eff}(E)$ is the effective pair $NN$ potential of the dibaryon-induced model, and  $W_{3BF}(E)$ is the effective three-body force induced by one-sigma exchange between the dibaryon and $\alpha$-core. The form of $V_{NN}^{\rm eff}(E)$ and $W_{3BF}(E)$ is described below.
The pair $NN$ potential $V_{NN}^{\rm eff}(E)$ acting in the three-body system is obtained from the pure two-nucleon effective dibaryon-induced potential $v_{NN}^{\rm
eff}({\cal E})$ (see Eq. (\ref{vnneff}) in the next Section), according to a general recipe for the transition from the two-particle system to the three-particle one, i.e., by replacing the two-body energy $\cal E$ with the difference $E-q^2/2\bar{m}$ in all expressions containing the energy dependence:
\begin{equation}
V_{NN}^{\rm eff}(E)=v_{NN}^{\rm eff}(E-q^2/(2\bar{m}))\delta(\bf{q}-\bf{q}'),
\label {Wnneff}
\end{equation}
where $q$ is the relative $\alpha-NN$ momentum and $\bar{m}=2m_Nm_{\alpha}/(2m_N+m_{\alpha})$ is the reduced mass of the $\alpha$-particle and the nucleon pair.

As the $N\alpha$ potential, we employed the quasilocal potential with the even-odd splitting and with the forbidden state $|\Phi _{N\alpha } \rangle$, which was proposed and used in Ref.~\cite{NP95} (the so-called MS potential).
This quasilocal potential includes the projector onto the forbidden $0S$ state with a large constant $\mu$ and can be written in the following  form:
\begin{equation}
V_{N\alpha }= V_{C}+V_{ls}({\bf l}{\bf s})+\mu |\Phi _{N\alpha }\rangle\langle\Phi _{N\alpha }|,
\label{vna}
\end{equation}
\begin{equation}
V_{C}= V^{C}_{1}(r)+P_{x}V^{C}_{2}(r),
\label{vnac}
\end{equation}
\begin{equation}
V_{sl}= V^{ls}_{1}(r)+P_{x}V^{ls}_{2}(r),
\label{vnasl}
\end{equation}
where $P_{x}$ is the Majorana operator which realizes the space
reflection $ {\bf r}\rightarrow -{\bf r} $, and the central
and spin-orbit parts of the potentials $V_{1}$ and $V_{2}$ are
determined via the potentials $V_{ev}$ and $V_{od}$
describing even and odd partial waves:
$V_{1,2}=\frac{1}{2}\left [ V_{ev}\pm V_{od}\right ]$.
The potentials $V_{ev}$ and $V_{od}$ having the Gaussian form
\begin{equation} V(r) = g \exp(-\varkappa^{2}r^{2})
\label{vnagauss}
\end{equation}
with the parameters given in Table~\ref{parna} reproduce quite well  the $S$-, $P$- and $D$-wave phase shifts in $N -\alpha$ scattering at energies from 0 to 20 MeV.

\begin{table}[h]
\centering
\caption{Parameter values for the MS $N\alpha$ interaction
potential (from Ref. \cite{NP95}).}
\medskip
\begin{tabular}{ccccc} \hline
& \multicolumn{2}{c}{even waves}& \multicolumn{2}{c}{odd waves}  \\
\cline  {2-5} & $g$(MeV) & $\varkappa $(fm$^{-1}$) & $g$(MeV) & $\varkappa
$(fm$^{-1}$) \\ \hline
$V_c$   & -65.58 & 0.6203 & -46.303 & 0.43216 \\
\hline
$V_{sl}$ & -12.169 & 0.8032 &  -15.931 & 0.62816  \\
\hline
\end{tabular}
\label{parna}
\end{table}

\subsection{Variational formalism for $\alpha+2N$ system with dibaryon-induced $NN$ interaction}

To find the energies and eigenfunctions for the $A=6$ nuclei within the cluster $\alpha+2N$ model, we used
the variational formalism on the non-orthogonal multiscale Gaussian basis which was described in detail in
a series of studies of the structure of the $A=6$ nuclei~\cite{voronchev82,NP84,NP86,NP95}.  Here we give
only the form of the basis functions used in this work and the corresponding notation for the quantum
numbers.

The Schr\"odinger equation for the wave function of the $\alpha+2N$ system with the effective energy-dependent Hamiltonian $H^{\rm tot}(E)$ was solved by the variational method using the Gaussian basis.
To solve this equation with the explicit energy dependence of
the effective Hamiltonian (\ref{htot}),
we used an iterative procedure with respect to the total energy E:
\[ H^{\rm tot}(E^{(n-1)})\Psi^{(n)}=E^{(n)}\Psi^{{\rm ex}(n)}.
 \]
 Such iterations can be shown to converge if the energy derivative of the effective interaction is negative. For bound states, i.e., for $E<0$, this condition is always valid. In our calculations, 5--10 iterations provided usually the accuracy of 5 decimal digits for the binding energy.

The total wave function  $\Psi^{JM_{J}}(\bm{r},\bm{\rho})$ of the three-body system with the total
angular momentum $J$ and its projection $M_J$ is expanded in a series of the  basis functions $\Phi^{(i)}_{\gamma n}$:
\begin{equation}
  \Psi^{JM_J}(\bm{r},\bm{\rho})=\sum_{\gamma}\sum_{n}
  C^{\gamma}_n \Phi_{\gamma n}(\bm{r},\bm{\rho}).
 \label{psi3}
 \end{equation}
 where $\gamma=\{\lambda,l,L,S\}$ and
 $C^{\gamma}_n$ are the unknown coefficients or linear
variational parameters. The six-dimensional basis functions $\Phi^{(i)}_{\gamma
n}$ are constructed from the Gaussian functions and the corresponding spin-angular
factors:
\begin{equation}
 \Phi_{\gamma n}(\bm{r},\bm{\rho})=N^\gamma_nr^{\lambda}\rho^{l}
 \exp\{-\alpha_{\gamma n}r^2 - \beta_{\gamma n}\rho^2\} \Omega^{JM_{J}}_\gamma (\hat{\bm{r}},\hat{\bm{\rho}}).
  \label{Phigamn}
 \end{equation}
The spin-angular part $\Omega^{JM_{J}}_\gamma (\hat{\bm{r}},\hat{\bm{\rho}})$ is taken in the following form:
\begin{eqnarray}
\label{3}
\Omega^{JM_{J}}_\gamma(\hat{\bm{r}},\hat{\bm{\rho}})=\sum_{M_{L}M_{S}}\langle LM_{L}SM_{S}|JM_{J}\rangle \nonumber \\ y^{LM_{L}}_{\lambda l}(\hat{\bm{r}},\hat{\bm{\rho}})\chi^{SM_{S}}(2,3),
\end{eqnarray}
where $y^{LM_{L}}_{\lambda l}(\hat{\bm{r}},\hat{\bm{\rho}})$ is the standard angular tensor and $\chi^{SM_{S}}(2,3)$ is the spin function of the nucleon pair.

The nonlinear parameters of the basis functions $\alpha_{\gamma n}$ and
$\beta_{\gamma n}$ are taken on the Chebyshev grid, which provides the
completeness of the basis and fast convergence of the variational
calculations~\cite{NP84,PLB83}:
\begin{eqnarray}
\alpha_{\gamma n}=\alpha_{\gamma
0}\tan^{a_\gamma}\left(\frac{\pi}{2}\frac{2n\!-\!1}{2N_{\alpha\gamma}}\right), \, \text {$n=1,...N_{\alpha\gamma}$,} \nonumber \\
\beta_{\gamma n}=\beta_{\gamma
0}\tan^{b_\gamma}\left(\frac{\pi}{2}\frac{2n\!-\!1}{2N_{\beta\gamma}}\right)), \, \text {$n=1,...N_{\beta\gamma}$,}
\label{Cheb}
\end{eqnarray}
where $N_{\alpha\gamma}$ ($N_{\beta\gamma}$) is the basis dimension for the variable
$r$ ($\rho$) for the channel with the quantum numbers $\gamma$, and the
parameters $a_\gamma$ ($b_\gamma$) are selected to be in the range 0.8--5
from the condition of optimality of the basis and stability of the calculation.

As was demonstrated earlier~\cite{NP84,NP86,NP95,Varga}, such a multidimensional  Gaussian basis is very flexible and can reproduce even quite complicated correlations in few-body systems.  An important advantage of the Gaussian basis in the few-body calculations is that it allows for a calculation of all the necessary matrix elements of the realistic Hamiltonian (including the spin-orbit and tensor forces) in the fully analytical form \cite{voronchev82,Varga}.
Another advantage of such a basis is that the form of the Gaussian functions is the same both in the coordinate and momentum representations. Thus the (normalized) basis functions
$\Phi_{{\gamma}n}^{(i)}({\bf p}_i,{\bf q}_i)$ in the momentum representation have the same form as in Eq.~(\ref{Phigamn}):
\begin{equation}
\Phi_{{\gamma}n}({\bf p},{\bf q})=
\tilde{N}^{\gamma}_np^{\lambda}q^l\,
\exp(-\tilde{\alpha}_np^2-\tilde{\beta}_nq^2)
\Omega^{JM_{J}}_\gamma(\hat{\bf p},\hat{\bf q})
\label{Phiq}
\end{equation}
where
\begin{equation}
\tilde{\alpha}_n=\frac{1}{4\alpha_n},\;\tilde{\beta}_n=\frac{1}{4\beta_n}.
\end{equation}
This feature of the Gaussian basis is especially significant when using the dibaryon-induced force model, since the effective pair $NN$ potentials and three-body forces explicitly depend on the momentum of the third particle.

\section{Dibaryon model for $NN$ and $3N$ forces}
\label{Dib-model}
\subsection{Description of the basic $NN$ channels in $A=6$ nuclei}
The dibaryon model proposed initially in Ref.~\cite{PIYAF} and developed further in Refs.~\cite{JPhys2001,KuInt} (where it was called ``the dressed-bag model'') suggests the following picture of the interaction between nucleons.
At a relatively long distance ($r_{NN} > 1$~fm) the nucleons interact by the conventional pion exchange. However, when the nucleons come closer to each other (to a distance $r_{NN} \lesssim 1$~fm), a compound state is formed: two nucleons fuse into a dibaryon state which is a six-quark bag dressed by a field of the light scalar $\sigma$-mesons. As a result of multiple transitions of two nucleons to the state of a dressed six-quark bag and vice versa, an effective $NN$ interaction arises which gives the main attraction between the nucleons at intermediate distances.

To describe such an interaction mechanism, it is natural to use a two-channel formalism which assumes that a system of two nucleons can be in two different states (channels): an external $NN$ channel and an internal six-quark (dibaryon) channel.
The full wave function of such a system consists of two components belonging to two Hilbert spaces of different nature. The external component of the wave function depends on the relative coordinate (or momentum) of two nucleons and their spins and isospins, while the internal one depends on the quark and gluon (or string) variables. Two independent Hamiltonians, i.e., the external one $h^{\rm ex}$ and
the internal one $h^{\rm in}$, are defined in these two spaces, respectively.

The external $NN$ Hamiltonian includes the kinetic energy $t$ and the peripheral part of the one-pion- and two-pion-exchange (OPE and TPE, respectively) interactions as well as the Coulomb interaction (for two protons):
\[h^{\rm ex}=t+\{v^{\rm OPE}+v^{\rm TPE}\}
 + v^{\rm Coul}.\]
Excluding the internal dibaryon channel from the two-channel Schr\"odinger equation, one
obtains the effective energy-dependent $NN$ Hamiltonian, which includes the resolvent of the
internal channel $g^{\rm in}=(E-h^{\rm in})^{-1}$ and the operators for transitions from the external channel to the internal one, $h^{\rm in,ex}$, and backwards, $h^{\rm ex,in}=(h^{\rm in,ex})^\dag$.

In the simplest version of the model \cite{KuInt}, a single-pole approximation for the dibaryon (internal) resolvent $g^{\rm in}$ was used:
 \begin{equation}
g^{\rm in}(E)=\sum_{\zeta}\int\frac{|\zeta,{\bf k}\rangle
\langle \zeta,{\bf k}| d^3k}{E-E_{\zeta }(k)}.
\label{resn}
\end{equation}
Here $|\zeta\rangle$ is the six-quark part of the wave function for the dressed bag with the definite quantum numbers of the orbital angular momentum $\lambda$, spin $S$ and total angular momentum $\vec{\cal J}=\vec{\lambda}+\vec{S}$. For simplicity, we will  denote the set of these quantum numbers by the one symbol $\zeta=(\lambda, S, \cal{J})$. Note that since in this version of the model only the symmetric six-quark state $[s^6]$ was included in the internal channel, the orbital angular momentum of the dressed dibaryon $\lambda$ is equal to that of the $\sigma$-meson.
The state $|{\bf k}\rangle$ represents the free wave of the $\sigma$-meson propagation.
The total energy $E_{\zeta }(k)$ of such a dressed state is
 \begin{equation}
 E_{\zeta }(k) = m_{D}+\varepsilon_{\sigma}(k),
 \label{eak}
\end{equation}
where
  \begin{equation}
 \varepsilon_{\sigma}(k) = k^2/2m_{D}+\omega_{\sigma}(k) \simeq
 m_{\sigma}+k^2/2\bar{m}_{\sigma},
 \label{epssig}
\end{equation}
 $\bar{m}_{\sigma}= m_{\sigma}m_{D}/(m_{\sigma}+m_{D})$,  and $\omega_{\sigma}(k)=\sqrt{m_{\sigma}^2+k^2}$ is the relativistic energy
 of the $\sigma$-meson, $m_{\sigma}$ and $m_{D}$ are the masses of the $\sigma$-meson
 and the $6q$ bag, respectively.

\begin{figure}[h]
\begin{center}
\includegraphics[width=0.9\columnwidth]{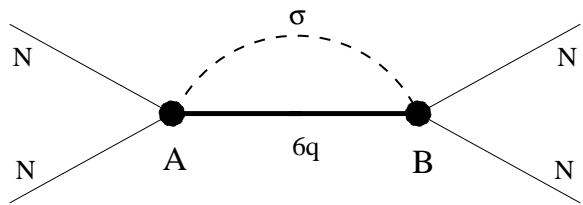}
\caption{Effective $NN$ interaction induced by production of an
intermediate dibaryon dressed by the $\sigma$ field.}
\label{vnqn}
\end{center}
\end{figure}

 The effective $NN$ interaction $w(E)$ resulting from the
coupling of the external $NN$ channel to the intermediate dibaryon state is illustrated by the graph in Fig. \ref{vnqn}.

To derive the effective interaction $w$ for the $NN$ channel in this approximation, one does not need to know the full internal Hamiltonian $h^{\rm in}$ of the dressed di\-baryon, as well as the full transition operator $h^{\rm ex, in}$.
One needs to know only how the transition operator $h^{\rm ex,in}$ acts on those dibaryon states $|\zeta,{\bf k}\rangle$, which are included into the resolvent (\ref{resn}).
The calculation of this quantity within the microscopical quark--meson model results in a sum of the factorized terms~\cite{KuInt}:
  \begin{equation}
  h^{\rm ex,in}|\zeta^{{\cal J}M},{\bf k}\rangle =
  \sum_{\lambda}|\varphi^{{\cal J}M}_{\lambda}\rangle B^{\cal J}_{\lambda}({\bf k}),
 \label{vert}
\end{equation}
where $\varphi^{{\cal J}M}_{\lambda} \in {\cal H}^{\rm ex}$ is the $NN$ transition
form factor and $ B^{\cal J}_{\lambda}({\bf k})$ is the vertex function dependent on the
$\sigma$-meson momentum. Here and below, where possible, to shorten the notation, we omit the index indicating the spin of the dibaryon.
The quantum numbers of the dibaryon $(\lambda, S,{\cal J})$ are naturally equal to the quantum numbers  of the $NN$ system from which the dibaryon is formed.

Thus, the effective interaction in the $NN$ channel $w(E)\equiv h^{\rm
ex,in}g^{\rm in}(E) h^{\rm in,ex}$ can be written as a sum of separable terms
in each partial wave:
 \begin{equation}
w(E)=\sum_{{\cal J} ,\lambda ,\lambda'}w^{\cal J}_{\lambda\lambda'}({\bf r},{\bf r}^{\prime},E),
 \label{nqn2}
 \end{equation}
 with
\begin{equation}
w^{\cal J}_{\lambda\lambda'}({\bf r},{\bf r'})= \sum_M \varphi^{{\cal J}M}_{\lambda}({\bf
r})\,\lambda^{\cal J}_{\lambda\lambda'}(E)\, {\varphi^{{\cal J}M}_{\lambda'}}^*({\bf r'}).
 \label{zlz}
 \end{equation}
The energy-dependent coupling constants
$\lambda^{\cal J}_{\lambda\lambda'}(E)$ appearing in Eq. (\ref{zlz}) are
directly calculated from the loop diagram shown in Fig.~\ref{vnqn}.
They are expressed in terms of the loop integral of the product of
two transition vertices $B$ and the convolution of two propagators
for the meson and six-quark bag with respect to the momentum $k$:
 \begin{equation}
\lambda^{\cal J}_{\lambda\lambda'}(E)=\int d{\bf k}
\frac{B_{\lambda}^{\cal J}({\bf k})\,{B_{\lambda'}^{\cal J}}^*({\bf k})}
{E-E_{\zeta}(k)}.
 \label{lamb}
 \end{equation}

The vertex form factors $B^{\cal J}_{\lambda}({\bf k})$ and the transition form factors
$\varphi^{{\cal J}M}_{\lambda}$
 have been calculated using the microscopic quark--meson
model~\cite{JPhys2001,KuInt}. Within this microscopic model, the transition form factors
$\varphi^{{\cal J}M}_{\lambda}$ are equal to $|2s\rangle$ and $|2d\rangle$ harmonic oscillator wave functions with a radius $r_0=\sqrt{\frac{2}{3}}b$, where $b=0.5$~fm. The explicit form of the vertex functions $B^{\cal J}_{\lambda}({\bf k})$ is not required to calculate the effective $NN$ potential.  However, these functions are necessary for calculating the three-body force due to two-sigma exchange (see below). From the microscopic quark model they can be derived in the Gaussian form~\cite{KuInt}:
\begin{equation}
B^{\cal J}_{\lambda}({\bf k})=B_0^{{\cal J}\lambda} \frac{{e}^{-b^2k^2}}{\sqrt{2\omega_{\sigma}(k)}},
 \label{Bk}
 \end{equation}
where $\bf k$ is the meson momentum and
\begin{equation}
b^2=\frac{5}{24}b_0^2,\qquad b_0=0.5\mbox{ fm}.
 \label{b0}
 \end{equation}
 The vertex constants $B_0^{{\cal J}\lambda}$ in Eq.(\ref{Bk}) must  satisfy Eq.(\ref{lamb}), i.e.:
 \begin{equation}
 \frac{1}{(2\pi)^3}\int d{\bf k}\frac{B_0^{{\cal J}\lambda}B_0^{{\cal J}\lambda'} {e}^{-2b^2k^2}}
 {(E\!-\!m_{\zeta}\!-\!\varepsilon_{\sigma}(k))\cdot 2\omega_{\sigma}(k)}
 = \lambda^{\cal J}_{\lambda\lambda'}(E),
 \label{adjust}
 \end{equation}
where $\lambda^{\cal J}_{\lambda\lambda'}(E)$ are the coupling constants found from the $NN$ phase shifts fitted within the model.

Thus, the dibaryon concept leads to an
{\em effective energy-dependent} $NN$ Hamiltonian in {\em the external channel}:
\begin{equation}
{\cal H}_{\rm eff}=t+v_{NN}^{\rm eff}(E),
\label {Hameff}
\end{equation}
where
\begin{equation}
v_{NN}^{\rm eff}(E)=w(E)+V_{\rm OPE}+V_{\rm TPE}+\lambda_{\rm orth} \Gamma .
\label {vnneff}
\end{equation}

 \begin{table*} [ht]
 \caption{Parameters of the dibaryon model found from the fit of $NN$
 scattering phase shifts up to $T_{\rm lab}=1$~GeV.}
 \begin{center}
\begin{tabular}{c|ccccccccc} \hline
    & $r_0,$ & $r_2,$ & $\lambda_{00},$ &
    $\lambda_{22},$ & $\lambda_{02},$ &
      $E_0,$ & $a,$ & $V^0_{\rm TPE},$ & $\beta^{-2},$\\
   & fm & fm & MeV & MeV & MeV & MeV & & MeV & fm$^{-2}$ \\    \hline
$^3S_1 - {}^3D_1$ & 0.41356 & 0.59423 & -328.55 & -15.65 & 44.06 &
693 & -0.05 & -4.0573 & 0.5301\\ \hline
$^1S_0$ & 0.430 & - & -328.9& - & - & 693 & -0.05 &
-8.803 & 0.6441\\ \hline
\end{tabular}
\end{center}
\label{parDBM}
\end{table*}

The effective $NN$ interaction $v_{NN}^{\rm eff}(E)$ includes
the peripheral OPE potential $V_{\rm OPE}$ with
a soft dipole cut-off with parameter $\Lambda=700$~MeV and
 a {\em peripheral} $2\pi$-exchange  contribution $V_{\rm TPE}$ which has been
 imitated in \cite{KuInt} by the form:
 \begin{equation}
      V_{\rm TPE}=V^0_{\rm TPE}\,(\beta r^2)^2\exp(-\beta r^2),
\label{twopi}
 \end{equation}
with the parameter values given in Table~\ref{parDBM}. From this Table one can conclude
that the strength of this potential is quite small (ca. 4--8~MeV).
However, this small
 contribution in the region $r \sim 2$~fm
is needed to reproduce exactly the effective-range parameters and
 the low-energy phase shifts.

The microscopic quark model also implies an orthogonality condition, which is provided by the term $\lambda_{\rm orth} \Gamma$ in the effective $NN$ interaction (\ref{vnneff}):
\begin{equation}
\lambda_{\rm orth} \Gamma =\lambda_{\rm orth}|0s\rangle\langle 0s|
\end{equation}
with $ \lambda_{\rm orth}\geqslant 10^6$~MeV and the same value of the oscillator radius $r_0=\sqrt{\frac{2}{3}}b$  as in the transition form factors  for the $s$-states.

Finally, we give the explicit formulas for the effective $NN$ interaction $w(E)$ in the lowest partial waves induced by the intermediate dressed dibaryon, which are
used in the present calculations of the $A=6$ nuclei:

 -- for the singlet $^1S_0$ channel:
\begin{equation}
w(E)= \lambda^0_{000}(E)|0s\rangle \langle 0s|,
\label {mods}
\end{equation}

 -- for the triplet coupled $^3S_1$--$^3D_1$ channels:
\begin{equation}
w(E)=\left (\!
\begin{tabular}{cc}
   $\lambda^1_{100}(E)|2s\rangle \langle 2s| \,$
 & $\lambda^1_{102}(E)|2s\rangle \langle 2d|$ \\
   $\lambda^1_{120}(E)|2d\rangle \langle 2s| \,$
 & $\lambda^1_{122}(E)|2d\rangle \langle 2d|$
\end{tabular} \! \right ),
\label{modt}
\end{equation}
where $|0s\rangle$, $|2s\rangle$ and $|2d\rangle$ denote the harmonic oscillator functions $|N\lambda\rangle$ with the number of quanta $N$ and orbital angular momentum $\lambda$.

In Ref. \cite{KuInt}, to reproduce the energy dependence of the constants
$\lambda^{\cal J}_{S\lambda\lambda'}$ derived from the above microscopic calculation (\ref{lamb}),
a Pade approximant [1,1] with two parameters $E_0$ and $a$ was used:
 \begin{equation}
 \lambda (E) =\lambda(0) \frac{E_0+aE}{E_0-E}.
 \label{pade}
 \end{equation}

\begin{figure}[h]
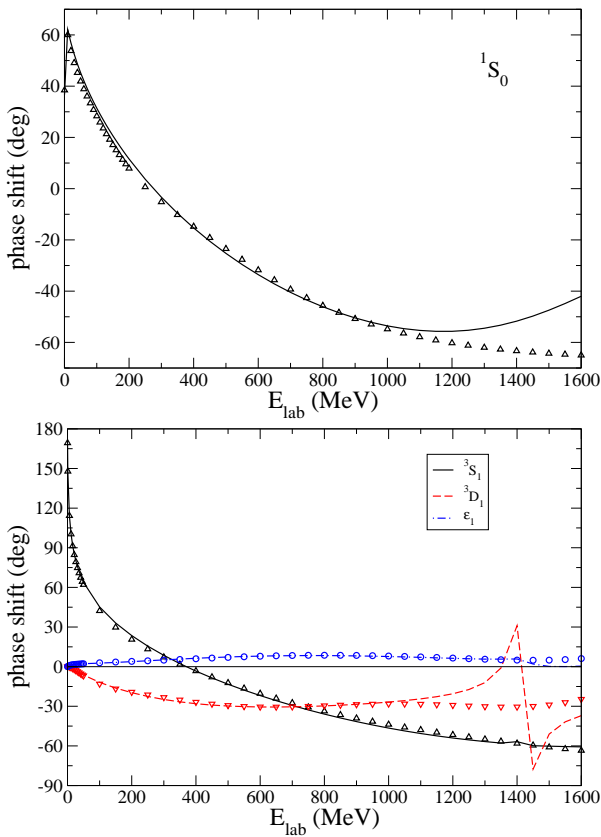

\begin{center}
\epsfig{file=fig3a.eps,width=0.45\textwidth}\\
\epsfig{file=fig3b.eps,width=0.45\textwidth}\\
\end{center}
% \setcaptionmargin{0mm} \onelinecaptionsfalse
%\captionstyle{flushleft}
\caption{The energy dependence of the $NN$ phase shifts obtained within the dibaryon-induced model with the parameter values from Table~\ref{parDBM} in comparison with the SAID partial-wave analysis data (solution SP07)~\cite{SAID} (open triangles and circles): top panel --- $^1S_0$ phase shifts,
bottom panel --- $^3S_1$ (solid curve) and $^3D_1$ (dashed curves) phase shifts, and the mixing angle $\varepsilon_1$ (dot-dashed curve).}
\label{phase}
\end{figure}

The above dibaryon-induced model for the $NN$ force (see Eqs. (\ref{vnneff}) and (\ref{twopi})--(\ref{pade})) was employed in Ref.~\cite{KuInt} to describe the lowest $NN$ partial phase shifts until $T_{\rm lab}=1$~GeV by adjusting the parameters $r_0$, $\lambda(0)$, $E_0$, $a$ and also the ``$2\pi$-exchange'' parameters $V^0_{\rm TPE}$ and $\beta$. The parameter
values found from that fit are given in Table~\ref{parDBM}.

The quality of predictions for the $NN$ phase shifts in the singlet $^1S_0$ and triplet $^3S_1$--$^3D_1$ partial channels is shown in Fig.~\ref{phase}.
As is clearly seen from the Figure,
the above model gives a very good description for the lowest partial phase shifts in the energy region from zero up to 1 GeV~\cite{KuInt}.
Agreement with experimental data for the $NN$ phase shifts and also for the static deuteron properties found within this force model, in general, is better and spans much broader energy interval than for the so-called realistic $NN$ potentials such as Nijmegen, Argonne, etc.
The weight of the internal (dibaryon) component
in the deuteron was obtained in Ref.~\cite{KuInt} to be $P_{\rm in}=3.66\%$.

It is this version of the dibaryon model (with the parameters from Table~\ref{parDBM}) that
is used in the present work to describe the $A=6$ nuclei
within the framework of the cluster $\alpha+2N$ model.

 Having obtained the solution $\Psi^{\rm ex}$ for the Schr\"odinger equation with the effective Hamiltonian (\ref{Hameff}), one can recover
 the excluded internal component $\Psi^{\rm in}$ of the total wave function unambiguously:
 \begin{equation}
 \Psi^{\rm in} =g^{\rm in}(E)h^{\rm in,ex}\Psi^{\rm ex}.
 \label{psiin}
 \end{equation}
Here, we give only the explicit expression for the weight of internal dibaryon component $P_{\rm in}$ in the bound-state (in our case, the deuteron) wave function.
The norm of the internal component (with the given $\cal J$)
is:
\begin{align}
\|\Psi^{\rm in}_{{\cal J}M}\|^2=\|\alpha^{{\cal J}M}\|^2
\sum_{\lambda\lambda'}
\langle\varphi^{{\cal J}M}_{\lambda}|\Psi^{\rm ex}\rangle
\langle \Psi^{\rm ex}|\varphi^{{\cal J}M}_{\lambda'}\rangle \nonumber \\
\times\underbrace{\int\frac{B^{\cal J}_{\lambda}({\bf k})
{B^{\cal J}_{\lambda'}}^*({\bf k})}{(E-E_{\zeta}({\bf
  k}))^2} d{\bf k}}_{I_{\lambda\lambda'}^{\cal J}}.
 \label{normin}
 \end{align}

It is easy to see from the comparison of Eqs. (\ref{lamb}) and
(\ref{normin}) that the integral $I_{\lambda\lambda'}^{\cal J}$ in Eq.(\ref{normin}) is
equal to the energy derivative (with an opposite sign) of the
coupling constant $\lambda_{\lambda\lambda'}^{\cal J}(E)$:
\[ I_{\lambda\lambda'}^{\cal J}=-\frac{{\rm d}\lambda^{\cal J}_{\lambda\lambda'}(E)}{{\rm d}E}, \]
and therefore:
\[ \|\Psi^{\rm in}\|^2 \sim -\frac{{\rm d}\lambda(E)}{{\rm d}E}, \]
i.e., the weight of the internal dibaryon state is proportional to the energy
derivative (with an opposite sign) of the coupling constant of the effective $NN$ interaction.
In other words, the stronger the energy dependence of the interaction in the $NN$ channel,
the larger the weight of the channel corresponding to the non-nucleonic degrees of freedom.

Assuming that the total bound-state wave function $\Psi$ must be normalized to
unity, and the external (nucleonic) part of the wave function $\Psi^{\rm ex}$
found from the effective Schr\"odinger equation also has the standard
normalization $\|\Psi^{\rm ex}\|=1 $, one obtains the weight of the internal dibaryon
component as follows:
\begin{equation}
P_{\rm in}=\frac{\|\Psi^{\rm in}\|^2}{1+\|\Psi^{\rm in}\|^2}.
\label{P_in2}
\end{equation}

\subsection{Three-body force in $3N$ and $\alpha+2N$ systems within the dibaryon model}

In a system of several nucleons, each pair of nucleons can form an intermediate
dibaryonic state. Therefore, to describe such a system, it is necessary to use the
multichannel formalism that includes one external (nucleon) channel and several
internal (dibaryon) channels in accordance with the number of available nucleon
pairs. In the internal channels, a new interaction between the dibaryon and
other nucleons is possible.

\subsubsection{Three-body force in $3N$ system}

\begin{figure}[h]
\begin{center}
%\noindent {\epsfig{file=fig3a.eps,width=0.27\textwidth}}
 \epsfig{file=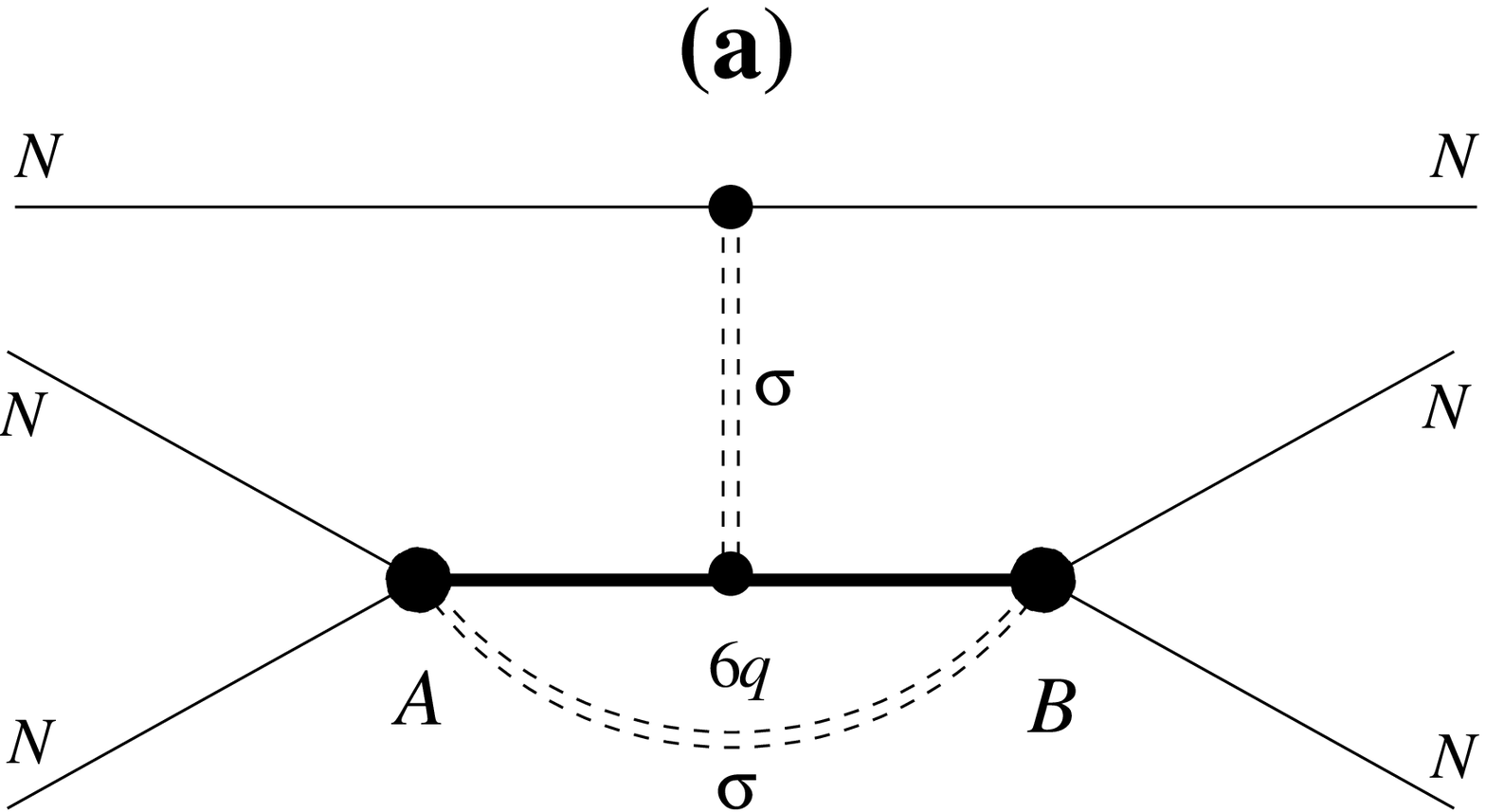,width=0.8\columnwidth}

 \bigskip
 \epsfig{file=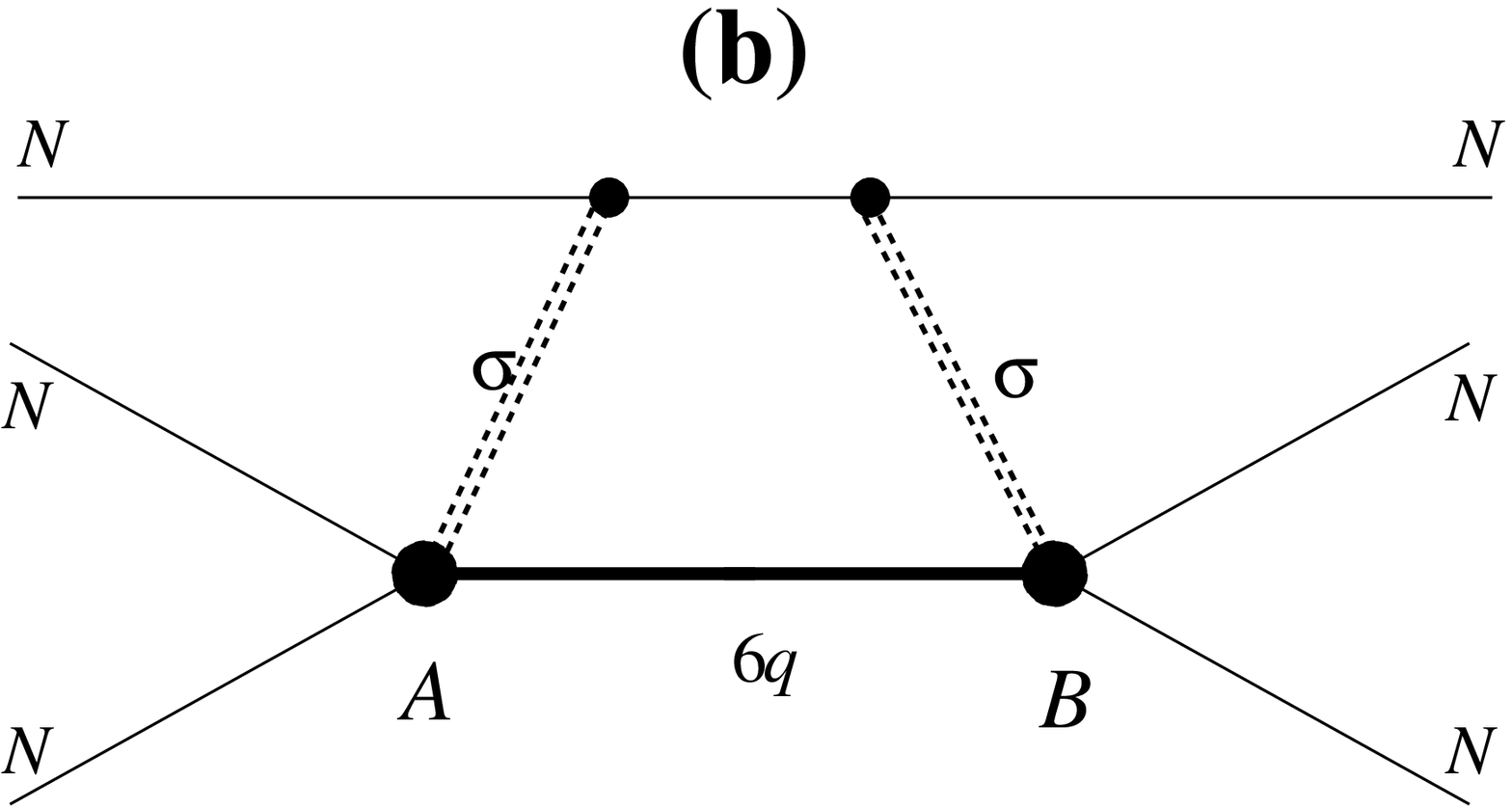,width=0.8\columnwidth}
\end{center}
% \setcaptionmargin{0mm} \onelinecaptionsfalse
%\captionstyle{flushleft}
\caption{The graphs corresponding to two types of the three-body force induced by $\sigma$ exchange between the dibaryon and the third nucleon.}
\label{fig3}
\end{figure}

In case of the three-nucleon system, one has a four-component space and thus one is dealing with a $(4\times 4)$-matrix Hamiltonian \cite{YAF3N,FBS2019}. Moreover, the dibaryon concept
inevitably leads to the appearance of a new three-body force (3BF) which arises mainly due
to the interaction of the third nucleon with the $\sigma$-meson field
surrounding the dressed dibaryon in each internal channel (see Fig. \ref{fig3}).

In this respect, the 3BF in the dibaryon model which is the immediate consequence of the basic two-nucleon force is in sharp contrast to the conventional 3BF like that of the Fujita--Miyazawa type, where the 3BF operator is not related directly to the traditional meson-exchange two-nucleon force and, moreover, includes constants (e.g., the cut-off parameters) other than those used for the initial two-nucleon force.
Therefore, in principle, one can add some 3BF contribution to the calculation with the conventional force model like RSC or AV18, but such a hybrid model should be considered as a fully phenomenological one.

The EFT approach where two- and three-nucleon forces are derived within the same paradigm is more consistent. It should also be noted that the role of three-particle forces turns out to be important in the halo-cluster EFT approach, e.g., in the description of the $alpha+N+N$ system.

In Refs. \cite{sys3n,YAF3N}, a dibaryon model for the $3N$ system was developed which included two types of three-nucleon forces caused by the $\sigma$-meson exchanges:
one-meson exchange (OSE) between the dressed bag and the third nucleon
(see Figs.~\ref{fig3}{\em a}) and the exchange by two $\sigma$-mesons (TSE),
where the third-nucleon propagator breaks the $\sigma$-loop of the two-body
force (Fig.~\ref{fig3}{\em b}).

These three-body forces can be included in the effective Hamiltonian for the external
$3N$ channel as some integral operators in the momentum space with the
factorized kernels:
 \begin{eqnarray}
  W^{\rm 3BF}_{(i)} ({\bf p}_i, {\bf p}'_i, {\bf q}_i, {\bf q}'_i;E)=
 \sum_{\small {\cal J}M,{\cal J}'M',\lambda ,\lambda'}\varphi^{{\cal J}M}_{\lambda}({\bf p}_i)\,\nonumber \\
 \times W^{{\cal JJ}'\lambda\lambda'}_{\rm 3BF}({\bf q}_i,{\bf q}'_i;E)
 \,\varphi^{{\cal J}'M'}_{\lambda'} ({\bf p}'_i),\;i=1,2,3,
 \label{3BF}
\end{eqnarray}
where ${\bf p}_i$ is the relative momentum of the nucleon pair $(jk)$ which generates the dibaryon,
${\bf q}_i$ is the momentum of the external relatively to this pair (spectator) nucleon $i$ and  $\varphi^{{\cal J}M}_{\lambda}({\bf p}_i)$ is the
transition form factor entering the effective $NN$ interaction (\ref{zlz}).
Here, the conventional numbering of particles in the three-body system is used:
$(ijk) \!= \!(123)$, (231), (312).
Thus, the matrix elements for the 3BF include only the overlap functions, and therefore the contribution of the 3BF is proportional to the weight of the internal $6qN$ components in the total $3N$ wave function.

 The operator for exchange by a scalar meson does not include any spin-
isospin variables, therefore in the case of the one-sigma exchange, the kernel of the 3BF operator is simplified and takes the form:
 \begin{align}
 W^{{\cal JJ}'\lambda\lambda'}_{\rm 3BF}({\bf q}_i,{\bf q}'_i;E)=
 \delta_{{\cal JJ}'} \int {d}{\bf k} \frac{B^{{\cal J}}_{\lambda} ({\bf k})}
  {E-E_{\zeta}(k)-q_i^2/2m} \nonumber \\
 \times V^{\rm OSE} ({\bf q}_i,{\bf q}'_i) \,
   \frac{B^{{\cal J}}_{\lambda'} ({\bf k})}{E-E_{\zeta}(k)-{q'}_i^2/2m},
\label{WOME}
\end{align}
 where
  \begin{equation}
  V^{\rm OSE} ({\bf q},{\bf q}') =
 \frac{-g^2_{\sigma NN}}{({\bf q}-{\bf q}')^2+m_\sigma^2}
 \label{vose}
 \end{equation}
 is the standard scalar-meson exchange potential.
 The integral over the $\sigma$-meson momentum $\bf k$ in Eq.~(\ref{WOME}) can be shown to
 be reduced to a difference in the values for the constant $\lambda(E-q^2/2m)$, so
 that the vertex functions $B(k)$ can be excluded from the formulas for the OSE 3BF
 matrix elements.
 The details of the calculations for such matrix elements are given in the Appendix of Ref.~\cite{YAF3N}.

The operator of the TSE interaction includes explicitly the vertex
functions $B(k)$ for the transitions ($NN \Longleftrightarrow
6q\!+\!\sigma$),  so that, these functions cannot be excluded
similarly to the case of the OSE interaction. In the calculations of the $3N$ nuclei \cite{sys3n,YAF3N}, the vertex functions (\ref{Bk}) with the parameter (\ref{b0}) were used. Here we do not give the formulas for the 3BF induced by TSE, since this contribution is not taken into account in the present calculations of the $A=6$ nuclei.

Table \ref{table3N} presents the results of calculations for  the ground states
of the $^3$H and $^3$He nuclei within the dibaryon-induced model for $NN$ and $3N$
forces (version designated as DBM(I) in Ref.~\cite{YAF3N}) in comparison with the  results obtained with the Argonne $NN$ potential AV18 and the
Urbana--Illinois three-body force UIX in Refs.~\cite{Pieper2001,Nogga2003}.

The AV18 interaction includes the charge symmetry breaking (CSB) by providing the $nn$ force, which is fitted to the experimental $nn$ scattering length different from the $pp$ one.
The electromagnetic part of the AV18 potential includes the one- and two-photon Coulomb terms, the Darwin--Foldy term, the vacuum polarization, the magnetic moment interactions, and a Coulomb term due to the neutron charge distribution. All these terms take into account the finite size of the nucleon charge distributions.
The calculations in Ref.~\cite{Nogga2003} have been done using the Faddeev equations for three identical fermions within the isospin formalism including total isospin states $T=1/2$ and 3/2. The small contribution of the $n$-$p$ mass difference to the ground-state energy was estimated perturbatively to be -7 keV for $^3$H and +7 keV for $^3$He.
\begin{table}[h!t]
% \setcaptionmargin{0mm} \onelinecaptionsfalse
%\captionstyle{flushleft}
\caption{Results of the $3N$ calculations from \cite{YAF3N} within the dibaryon model for
two- and three-body forces (DBM(I)).}
\begin{center}
\tabcolsep=0.25em
\begin{tabular}{*{8}{c}}  \hline
  Model&$E$&$P_D$&$P_{S'}$,&$P_{\rm in}$,
  & \multicolumn{3}{c}{Contributions to $H$}\\
  \cline{6-8}
        &MeV&\%&\%& \%                 &$T$& $T\!\!+\!\!V^{2N}$&$V^{3N}$ \\
 \hline
 \multicolumn{8}{c}{ $^3$H, $E_{\rm exp}=-8.482$~MeV}\\ \hline
 DBM(I)& --8.482 & 6.87 & 0.67 & 10.99& 112.8 & --1.33 & --7.15 \\
  AV18 + UIX& --8.476& 9.3 & 1.05 &  --   & 51.3 & --7.27 & --1.19\\
 \hline
\multicolumn{8}{c}{ $^3$He, $E_{\rm exp}=-7.718$~MeV}\\ \hline
 DBM(I)& --7.772  & 6.85 & 0.74 & 10.80 & 110.2 & --0.90 & --6.88\\
  AV18 + UIX& --7.746 & 9.25 & 1.24 & --     & 50.2 & --6.54 & --1.17\\
 \hline
\end{tabular}
\end{center}
\label{table3N}
\end{table}

 The results presented in Table~\ref{table3N} for the dibaryon model have been also obtained within the isospin formalism for three identical fermions with the use of the effective pair $NN$
potential (\ref{Hameff})--(\ref{pade}) with the parameters from
Table~\ref{parDBM} and the OSE and TSE 3BF with the following parameters:
\begin{equation}
g_{\sigma NN}=9.577, \, m_\sigma=400\,{\rm MeV},\, m_D=2240\,{\rm MeV}
\label{const3BF}
\end{equation}
The value of the $\sigma NN$ coupling constant in
the $^3$H calculations has been chosen to reproduce the exact binding
energy of the $^3$H nucleus.

 The dibaryon model with the same parameters was used to calculate the $^3$He nucleus in~\cite{YAF3N}. In the $^3$He nucleus, treated within the framework of the dibaryon model, in addition to the Coulomb interaction between protons, a new Coulomb three-particle force appears due to the Coulomb interaction between the proton and the charged dibaryon. When calculating the Coulomb interactions in~\cite{YAF3N}, the finite size of the charge distributions of the proton and dibaryon was taken into account.
So, the calculations for $^3$He have been
carried out {\em without any free parameters}.

 In Table~\ref{table3N} the following characteristics are given: the bound-state energy $E$, the weight $P_{\rm in}$ of the internal dibaryon channel, the weight $P_D$ of the $D$-wave component in the total $3N$ wave function as well as the weight $P_{S'}$ of the mixed-symmetry $S'$ component (only for the $3N$ channel); the average individual contributions from the kinetic energy $T$, the two-body interactions $V^{(2N)}$ plus the kinetic energy $T$, and the 3BF $V^{(3N)}$ due to OSE and TSE in the total Hamiltonian expectation. The contributions $V^{(2N)}$ and $V^{(3N)}$ for the AV18 + UIX model are taken from \cite{Pieper2001}.

Note that the TSE contribution for the $^3$H and $^3$He nuclei is substantial: it is about 40\% of the OSE contribution.
However, our calculations showed that the TSE contribution can be simulated by the OSE contribution with the increased coupling constant $g_{\sigma NN}$. Such a simplified model without the TSE-induced interaction when the coupling constant $g_{\sigma NN}$ is increased from 9.577 to 14.259 reproduces the basic properties of the ground states for the $^3$H and $^3$He nuclei just as well as the original model that takes into account both types of three-nucleon forces (OSE and TSE). Similarly to this, in our present calculations for the $A=6$ nuclei we fit $g_{\sigma NN}$ constant to account effectively for both OSE and TSE contributions.

\subsubsection{Three-body force in the $\alpha+2N$ system}

In the $\alpha+2N$ system, there is one pair of nucleons, therefore, within the framework of the dibaryon concept, the system must be described as two-channel one with one external and one internal channel.
In the external channel, the system consists of two nucleons interacting with each other and with the $\alpha$-particle, and in the internal channel one has a dressed $6q$ bag (dibaryon) interacting with the $\alpha$-particle.
This interaction between the dibaryon and the $\alpha$-particle in the internal channel, $V_{D\alpha}^{\rm in}$, leads to a three-particle force in the external channel, acting along the Jacobi coordinate of the $\alpha$-particle $\rho$ relative to the center of mass for the nucleon pair.

In the present work, as such a dibaryon-$\alpha$ potential in the internal channel, we use the folding potential obtained by averaging the OSE potential (\ref{vose}) between the $6q$ bag and the nucleon from the $\alpha$-cluster over the single-particle nucleon density of the $\alpha$-particle $\hat{\rho}(r_N)$ \cite{Tanihata92}:
\begin{equation}
\hat{\rho}(r_N)=4N_\rho \exp(-(r_N/r_\alpha)^2),\, \int \hat{\rho}(r_N) d^3r_N =1,
\label{adensity}
\end{equation}
where $r_N$ is the single-nucleon coordinate with respect to the center of mass of the $\alpha$-particle, $N_\rho$ is the normalization constant and $r_\alpha=1.325$~fm. It is easy to show that averaging the Yukawa potential (\ref{vose}) with the density (\ref{adensity}) results in the scalar-meson exchange potential (in the internal channel) with the standard exponential cutoff:
  \begin{equation}
  V_{D\alpha}^{\rm OSE}({\bf q},{\bf q}') =
 -g^2_{\sigma NN}\frac{\exp(-({\bf q}-{\bf q}')^2/\Lambda^2)}{({\bf q}-{\bf q}')^2+m_\sigma^2},
 \label{D-alpha}
 \end{equation}
where the cutoff parameter $\Lambda=2/r_\alpha$.

Thus, the three-body force in the $2N+\alpha$ system can be determined by the formula (\ref{WOME}) with a replacement of the $\sigma$-meson exchange potential (\ref{vose}) by the cutoff potential (\ref{D-alpha}):
 \begin{eqnarray}
 W_{\rm 3BF}^{{\cal J}\lambda\lambda'} ({\bf q},{\bf q'};E) =
  \int {d}{\bf k} \frac{B^{\cal J}_{\lambda} ({\bf k})}
  {E-E_{\zeta}(k)-q^2/2\bar{m}}\nonumber \\
  \times V_{D\alpha}^{\rm OSE} ({\bf q},{\bf q}') \,
   \frac{B^{\cal J}_{\lambda'} ({\bf k})}
  {E-E_{\zeta}(k)-{q'}^2/2\bar{m}}.
\label{W3BF}
\end{eqnarray}
where $\bar{m}=m_Dm_{\alpha}/(m_D+m_{\alpha})$, $m_\alpha$ is the mass of the
 $\alpha$-particle, and $m_D$ is the mass of the dibaryon.

As in case of the three-nucleon system, the integral in Eq.~(\ref{W3BF}) over the meson momentum $\bf k$ reduces to the difference of the coupling constants:
\begin{eqnarray}
 \int \frac{B^{\cal J}_{\lambda'}({\bf k})\,B^{\cal J}_{\lambda}({\bf k})}
 {(E\!-\!E_\zeta(k)\!-\!\frac{q^2}{2\bar{m}})
 (E\!-\!E_\zeta(k)\!-\!\frac{{q'}^2}{2\bar{m}})}\,
 {d}{\bf k} \nonumber \\
 = \frac{\lambda^{\cal J}_{\lambda'\lambda}(E\!-\!{q'}^2/2\bar{m})
 -\lambda^{\cal J}_{\lambda'\lambda}(E\!-\!q^2/2\bar{m})}
 {{q'}^2-{q}^2} \nonumber \\
 \equiv \Delta\lambda^{\cal J}_{\lambda'\lambda} (q',q,E).
\label{repl}
 \end{eqnarray}
For the
energy dependence $\lambda(E)$ taken according to Eq.~(\ref{pade}),
this difference is
\begin{equation} \Delta\lambda (q',q,E) =
\lambda(0)\!E_0(1+a)\frac{1}{E\!-\!E_0\!-\!\frac{q^2}{2\bar{m}}}\,
\frac{1}{E\!-\!E_0\!-\!\frac{{q'}^2}{2\bar{m}}}.
\label{ddlamb}
\end{equation}
Thus, the three-particle interaction in the $2N+\alpha$ system (in the external channel)
due to the $\sigma$-exchange averaged over the $\alpha$-particle density in
the internal channel takes the form:
\begin{eqnarray}
W_{\rm 3BF}^{{\cal J}\lambda\lambda'} ({\bf q},{\bf q'};E) =
 \lambda^{\cal J}_{\lambda'\lambda}(0)\,E_0(1+a)
  \frac{1}{E\!-\!E_0\!-\!\frac{q^2}{2\bar {m}}}  \nonumber \\
  \times \frac{-g^2_{\sigma NN}\exp(-({\bf q}-{\bf q}')^2/\Lambda^2)}{({\bf q}-{\bf q}')^2+m_{\sigma}^2}\,
  \frac{1}{E\!-\!E_0\!-\!\frac{{q'}^2}{2\bar {m}}}.
\label{WOSE}
\end{eqnarray}
In the present calculations for the $A=6$ nuclei, we used just this form for
the three-body force with the parameters $m_D$ and $m_\sigma$ from Eq.~(\ref{const3BF}).
We consider the coupling constant $g_{\sigma NN}$ as the only adjusted parameter.
However, it turned out that the value $g_{\sigma NN}=9.577$ adopted in the $^3$H calculations reproduces the $^6$Li binding energy very well.
Therefore, we decided not to change this value.

\section{Properties of $A=6$ nuclei within the dibaryon-induced model for $NN$ and $NN\alpha$ forces}

We performed three series of the complete variational
calculations within the $\alpha+2N$ cluster model using the same $\alpha N$  potentials which
include the odd-even splitting and the Pauli-projection operator (see Eqs. (\ref{vna})--(\ref{vnagauss}) and Table~\ref{parna}) but with different $NN$-interaction models:\\
 (i) the conventional $NN$ potential from the RSC model;\\
 (ii) the dibaryon-induced $NN$ potential (see Eqs. (\ref{Wnneff}), (\ref{vnneff}), (\ref{twopi})--(\ref{pade}) and Table \ref{parDBM} for the parameter values);\\
 (iii) the same dibaryon-induced $NN$ potential as in point (ii) but with added
OSE three-body (\ref{WOSE}) with the constant $g_{\sigma NN}$ chosen to reproduce the binding energy of $^6$Li.\\

We should emphasize that using the above fixed parameter value
$g_{\sigma NN}$, the calculation for $^6$He nucleus was done without any new fit parameters.

The Table~\ref{basis} presents the basis configurations
(the dimension and quantum numbers) used in all versions of the calculations
for the ground states of the $^{6}$He and $^{6}$Li nuclei.
\begin{table}[h!]
\caption { The dimension and quantum numbers of the basis for the separate spin-orbit channels $N_{\alpha}$ and $N_{\beta}$ included in the calculations of the $^{6}$He and $^{6}$Li ground states}
%\vspace{-6mm}
\begin{center}
\begin{tabular}{ccccc}
\hline
\rule{0mm}{5mm} Nuclei & $J^{\pi}T$ &  $\gamma=\lambda lLS$  & ${\cal J}=\lambda+S$ & $N_{\alpha\gamma}\times N_{\beta\gamma}$\\
\hline
\multirow{4}{*}{$^{6}$He} & \multirow{4}{*}{$0^{+}1$}
     & 0000  &  0 & 12$\times$10\\
  &  &  1111 &  1 & 10$\times$10 \\
  &  &  2200 &  2 & 10$\times$10 \\
  &  &  3311 &  3 & 10$\times$10 \\
\hline
\multirow{5}{*}{$^{6}$Li} & \multirow{5}{*}{$1^{+}0$}
     & 0001 &  1 & 15$\times$7\\
  &  & 2021 &  1 & 15$\times$7 \\
  &  & 1110 &  1 &  9$\times$7 \\
  &  & 2201 &  1,2,3 & 9$\times$7 \\
  &  & 2211 &  1,2,3 & 9$\times$7 \\
  &  & 2221 &  1,2,3 & 9$\times$7 \\
  &  & 0221 &  1 & 9$\times$7 \\\hline
\end{tabular}
\end{center}
\label{basis}
\end{table}

\subsection{Binding energies, weights for various spin-angular components and static properties of $A=6$ nuclei}
\begin{table*}[h]
\caption  {Properties of the ground states of the $^{6}$He and $^{6}$Li nuclei calculated with the RSC $NN$ potential and with the dibaryon-induced $NN$ potential without the three-body force (DM) and with the three-body force(DM+3BF).}

\begin{center}
%\begin{tabular}{|c|c|c|c|c|c|c|c|c|}
\begin{tabular}{|c|c|ccccc|cc|}
\hline
%\rule{0mm}{5mm}
\multirow{2}{*}{Nuclei} & \multirow{2}{*}{Model}  & \multirow{2}{*}{$E$, MeV} &

 \multirow{2}{*}{ $T$} & \multirow{2}{*}{ $V_{NN}$}  & \multirow{2}{*}{ $W^{\rm 3BF}$} &
 \multirow{2}{*}{$P_{\rm in}$,\%}&
   \multicolumn{2}{c|}{Configuration}\\
   \cline{8-9}
  &&& &&& & $\gamma=\{\lambda lLS\}$ & $P_\gamma$,\%\\
\hline
\multirow{13}{*}{$^{6}$He} &
\multirow{4}{*}{RSC}  & \multirow{4}{*}{-0.2761}  &
\multirow{4}{*}{26.084} & \multirow{4}{*}{-6.866} & \multirow{4}{*}{0}  &
  \multirow{4}{*}{0}  &
0000 & 88.091\\
&&&&&&&1111 & 9.872 \\
&&&&&&&2200 & 1.513\\
&&&&&&&3311 & 0.524\\
\cline{2-9}
& \multirow{4}{*}{DM}  & \multirow{4}{*}{-0.0055}  &
\multirow{4}{*}{33.322} & \multirow{4}{*}{-17.238}& \multirow{4}{*}{0}  &
 \multirow{4}{*}{2.32}  &
0000 & 88.954\\
&&&&&&&1111 & 8.901 \\
&&&&&&&2200 & 1.610\\
&&&&&&&3311 & 0.535\\
\cline{2-9}
& \multirow{4}{*}{DM+3BF}  & \multirow{4}{*}{-0.723}&
\multirow{4}{*}{43.235} & \multirow{4}{*}{-23.066}  & \multirow{4}{*}{-1.053}  &
 \multirow{4}{*}{2.99}    &
      0000 & 88.873\\
&&&&&&&1111 & 9.280 \\
&&&&&&&2200 & 1.380\\
&&&&&&&3311 & 0.467\\
\cline{2-9}

& {Experiment}  & {-0.975}  & && &   &  & \\

\hline

\multirow{17}{*}{$^{6}$Li} &
\multirow{7}{*}{RSC}  & \multirow{7}{*}{-3.225}  &
\multirow{7}{*}{41.306} & \multirow{7}{*}{-28.225}  & \multirow{7}{*}{0}  &
 \multirow{7}{*}{0}  &
0001 & 86.815\\
&&&&&&&2021 & 7.009\\
&&&&&&&1110 & 2.157\\
&&&&&&&2201 & 0.473\\
&&&&&&&2211 & 0.284\\
&&&&&&&2221 & 0.028\\
&&&&&&&0221 & 0.233\\
\cline{2-9}
& \multirow{7}{*}{DM}  & \multirow{7}{*}{-2.691}  &
\multirow{7}{*}{53.420} & \multirow{7}{*}{-42.671}  & \multirow{7}{*}{0}  &
  \multirow{7}{*}{4.34}  &
0001 & 91.353\\
&&&&&&&2021 & 5.808 \\
&&&&&&&1110 & 1.834\\
&&&&&&&2201 & 0.483\\
&&&&&&&2211 & 0.278\\
&&&&&&&2221 & 0.028\\
&&&&&&&0221 & 0.216\\
\cline{2-9}
& \multirow{7}{*}{DM+3BF}  & \multirow{7}{*}{-3.678}  &
\multirow{7}{*}{63.896} & \multirow{7}{*}{-49.138}  & \multirow{7}{*}{-1.389}  &
 \multirow{7}{*}{4.90}  &
0001 & 90.58\\
&&&&&&&2021 & 6.056 \\
&&&&&&&1110 & 2.270\\
&&&&&&&2201 & 0.453\\
&&&&&&&2211 & 0.321\\
&&&&&&&2221 & 0.045\\
&&&&&&&0221 & 0.277\\
\cline{2-9}
& {Experiment}  & {-3.70}  &&&&    &  & \\
\hline
\end{tabular}

\end{center}
\label{properties}
\end{table*}

\begin{table*}[h!]
\caption  {Static observables of  $^{6}$He and $^{6}$Li nuclei}

\begin{center}
\begin{tabular}{|c|c|cccccc|}
\hline
Nuclei & Model &$\sqrt{r^{2}}$, fm & $\sqrt{\rho^{2}}$, fm & $\sqrt{r_c^2}$, fm & $\sqrt{r_m^2}$, fm& $Q$,mb& $\mu$, \\ \hline
\multirow{4}{*}{$^{6}$Li} &
{RSC}    & 3.36  & 3.99  & 2.56  &  2.43         & 5.49  & 0.830  \\ \cline{2-8}
& DM     & 3.497  &  4.45 & 2.74  &  2.62         & 4.96  & 0.839  \\ \cline{2-8}
& DM+3BF & 3.323 & 3.70  & 2.46  &  2.32         & 4.59  & 0.835  \\ \cline{2-8}
& Experiment &   &       & 2.589(39)  & 2.42(4) & -0.818(17)  & 0.8220473(6)  \\ \hline
\multirow{4}{*}{$^{6}$He} &
{RSC}    & 4.555  & 3.877  & 2.085  & 2.55  &   &   \\ \cline{2-8}
& DM     & 4.837  & 4.468  & 2.212  & 2.80  &   &   \\ \cline{2-8}
& DM+3BF & 4.337  & 3.647  & 2.038  & 2.44  &   &   \\ \cline{2-8}
& Experiment &    &        & 2.054(14)  & 2.48(3)  &   &   \\ \hline
\end{tabular}
\end{center}
\label{observ}
\end{table*}

The results for all three series of calculations (i), (ii) and (iii) for the $^{6}$He
and $^{6}$Li nuclei are summarized in Tables \ref{properties} and \ref{observ}. In Table \ref{properties} we present the
binding energies $E$, the contributions of the separate terms to the Hamiltonian:
the kinetic energy $T$, the two-nucleon interaction $V_{NN}$ and the three-body force $W^{\rm
3BF}$, and also the weight of the dibaryon state $P_{\rm in}$ and the weights of different
spin-orbit components $P_\gamma$ in the total three-cluster wave function.
From Tables \ref{properties} and \ref{observ} one can conclude the following:

1. For the conventional $NN$ potential (RSC) the ground states are seriously underbound both for the $^{6}$He and $^{6}$Li nuclei. This is well known from Refs. \cite{NP84,NP86,NP95} and is also in agreement with Ref.~\cite{Horiuchi} where authors tested a few different $NN$ potentials in the similar $\alpha+2N$ calculations.

2. Replacement of the conventional pairwise $NN$ interaction with the dibaryon-induced interaction (keeping the $\alpha+N$ interaction to be the same) results in even less binding for both $^{6}$He and $^{6}$Li. This decrease of the binding energies is due to the strong enhancement of the average kinetic energy in the dibaryon model.

3. Inclusion of the three-body force due to $\sigma$-meson exchange between the dibaryon and the nucleons in the $\alpha$-core results in an increase of the binding energies and a significant improvement in agreement between the calculated binding energies and their experimental values.

4. The weights of the dibaryon components in the $n-n$ and $n-p$ subsystems are also increased when the three-particle force is included in the calculation.

5. When replacing the traditional model for the $NN$ force with the dibaryon-induced model, the average kinetic energy in both nuclei is increased remarkably.

6. General agreement with experimental data for the rms radii of the charge and matter distributions calculated within the DM+3BF model (see  Table~\ref{observ}) gets also better than that for the case without 3BF.

7. However, the results for the rms radii of the charge and matter distributions show, unfortunately, some opposite trends, viz., the results within the dibaryon-induced model for $^{6}$Li get worse while those for $^{6}$He get better as compared with the RSC results.
The worsening of the results for the rms radii for $^{6}$Li can be explained by ignoring some higher spin-angular components in the variational basis used in our calculations. The impact of these higher components in the $^{6}$Li wave functions should be much larger than in the $^{6}$He case.

From the above results one can conclude that the binding energy of the two-nucleon cluster in the field of the $\alpha$-core becomes higher, i.e., the two-nucleon coupling in the field of the core gets stronger. This result is especially interesting for the two-neutron system: for the traditional $NN$ force the $nn$ pair is unbound in case of an isolated $nn$ system and is very weakly bound in the $\alpha+2n$ system.
In contrast to this, when replacing the traditional $NN$ force with the dibaryon-induced one plus three-body OSE interaction, the effective $nn$ coupling gets much stronger and resembles the Cooper pairing in solids. (It is worth to remind the reader that the Cooper pairing between two electrons in solids appears due to the interaction of the electron pair with the lattice, i.e., with the third body, quite similarly to our case).
It is informative to note that the traditional $3N$-force mechanism of the Fujita--Miyazawa type does not work in the $\alpha+NN$ system due to the scalar-isoscalar nature of the $\alpha$-cluster.  The contact three-body force also makes a very low contribution in the $\alpha+2N$ cluster system because of the large interparticle distance in $^{6}$He.  So, this novel dibaryon-induced mechanism for the additional coupling of the neutron pair in the field of the $\alpha$-core can be considered as a good candidate for neutron pairing in other even-even nuclei and nuclear matter.

\subsection{Nucleon momentum distributions}

In the numerous experiments done for the last decades, a considerable enhancement in the nucleon momentum distributions in light and heavy nuclei at high transferred momenta $p_{m}\gtrsim 300 \,{\rm MeV}/c$ has been found, as compared to the predictions of the traditional $NN$-force models~\cite{Hen,23,24}.
The most impressive enhancement has been established in the experiments on scattering of high-energy electrons off nuclei done in JLab~\cite{J-Lab1,J-Lab2,J-Lab3,J-Lab4}.
Therefore, it would be very instructive to test the force models employed for the nuclear structure calculations from the point of view of the nucleon momentum distributions, especially at $p_{m}\gtrsim 300\,{\rm MeV}/c$.

We consider the momentum distributions of the relative motion of the valence nucleons in nuclei with two valence nucleons within the three-cluster model. We compare these distributions calculated with the traditional $NN$-force model and the dibaryon-induced model for the $NN$ and three-body $NN\alpha$ interactions.
These distributions ${\hat\rho}(p)$ are determined by the moduli squared of the components of the wave function in the momentum representation integrated over the momentum $\bf q$  and averaged over the angles of the momentum $\bf p$. We consider only the sums of components with the fixed value of the orbital angular momentum of a nucleon pair $\lambda$, namely with $\lambda=0$ ($S$-wave) and $\lambda=2$ ($D$-wave):
\begin{equation}
\hat{\rho}_{\lambda}(p)=\sum_{lLS}\int |\Psi_{\lambda lLS}(\bm p,\bm q)|^2 d{\bm q} \frac{d\hat{\bm{p}}}{4\pi}
\label{mom-dist}
\end{equation}

\begin{figure}[h!]
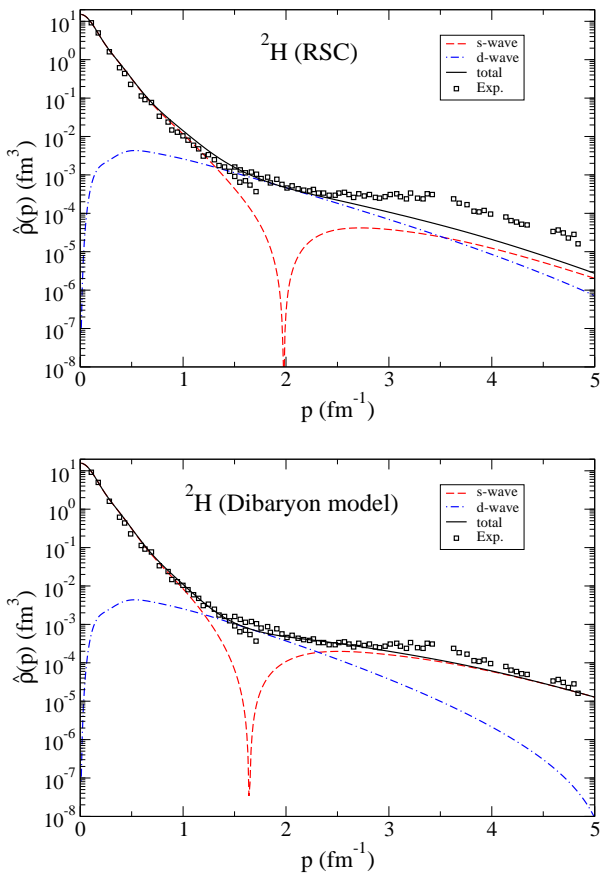

\begin{center}
\includegraphics[width=0.45\textwidth]{fig5a.eps}

\bigskip
\includegraphics[width=0.45\textwidth]{fig5b.eps}
\caption{The momentum distribution of the $np$
relative motion in the deuteron for the traditional RSC $NN$ potential (top panel) and for the dibaryon-induced model (bottom panel): $S$-wave contribution --- dashed curves, $D$-wave contribution --- dot-dashed curves, and their sum --- solid curves. Experimental data (squares) are taken from Ref. \cite{Blomq}.}
\end{center}
\label{mom_deut}
\end{figure}

First, we compare the momentum distributions of the $np$ relative motion in the deuteron calculated with the RSC $NN$ potential and the dibaryon-induced model. In
case of the deuteron, the corresponding distributions $\rho_\lambda$ are simply the squares
of the $S$- and $D$-wave components of the wave function. In Fig.~\ref{mom_deut} these
distributions are shown in comparison with the experimental data \cite{Blomq}.

As can be seen from the Figure, the dibaryon-induced model leads to good agreement
with the experiment for the momentum distribution up to $p=5$ fm$^{-1}$, in contrast to the
traditional RSC model for the $NN$ interaction. It is well known that the $S$-wave contribution
to the $np$ momentum distribution in the deuteron has a dip at $p\sim 2$~fm$^{-1}$
which is filled up by the $D$-wave component of the deuteron. When replacing the RSC
potential with the dibaryon-induced interaction, the dip in $S$-wave contribution
shifts towards lower momenta and the values of the $S$-wave distribution at $p
>2$~fm$^{-1}$ increases by almost an order of magnitude.

\begin{figure}[h!]
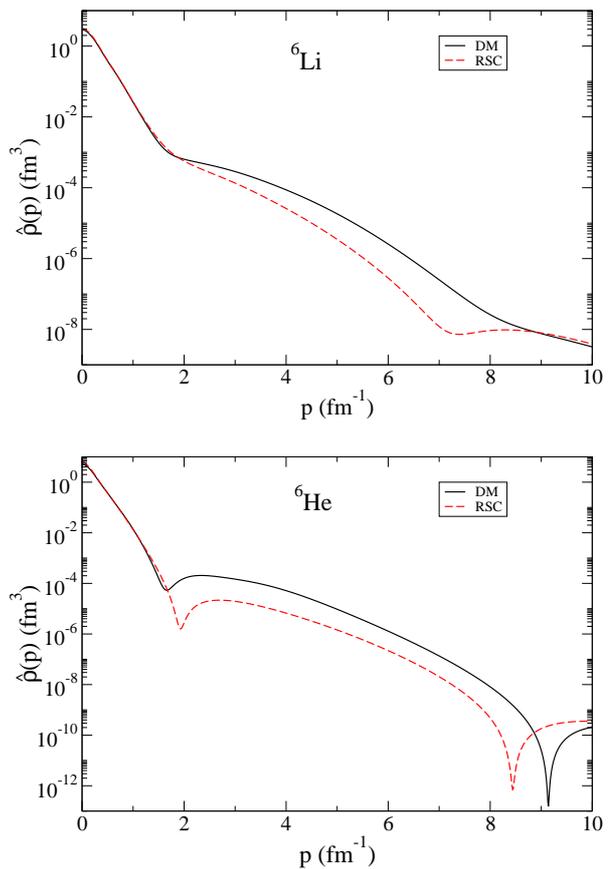

\begin{center}
\includegraphics[width=0.45\textwidth]{fig6a.eps}

\bigskip
\includegraphics[width=0.45\textwidth]{fig6b.eps}
\caption{The momentum distributions of the valence nucleons in $^6$Li
and $^6$He calculated within the dibaryon-induced model (solid curves) and the RSC potential (dashed curves).}
\end{center}
\label{LiHe-dist}
\end{figure}

A similar picture is observed for the momentum distributions of the valence nucleons in the $^6$Li and $^6$He nuclei showed in Fig.~\ref{LiHe-dist}.
It should be stressed that all results in this subsection, referred to as the dibaryon model results and denoted in Figs.~6--8 as DM, have been obtained within the dibaryon-induced model with taking the dibaryon-induced 3BF into account.

It can be seen from Fig.~\ref{LiHe-dist} that in the broad momentum range $p= 1.7$--$9$~fm$^{-1}$, the momentum distributions for the dibaryon model are several times higher than those for the traditional RSC potential. It means that the replacement of the traditional $NN$ potential with the dibaryon-induced interaction leads to a strong increase of the short-range $NN$ correlations.

To estimate the magnitude of the high-momentum components in the momentum distributions, we calculated the probability that the relative momentum of two nucleons $(p\geqslant p_{0})$ in the components of the wave functions for the
$^2$H, $^6$Li, and $^6$He nuclei found with two interaction models, i.e., the RSC and dibaryon-induced ones:
\begin{equation}
W_{\lambda}(p_{0})=\int_{p_0}^\infty \hat{\rho}_\lambda (p) p^2dp
\label{wp}
\end{equation}
These probabilities determine the magnitude of the short-range $NN$ correlations in the respective nuclei.

\begin{figure}[h]
\begin{center}
\includegraphics[width=0.45\textwidth]{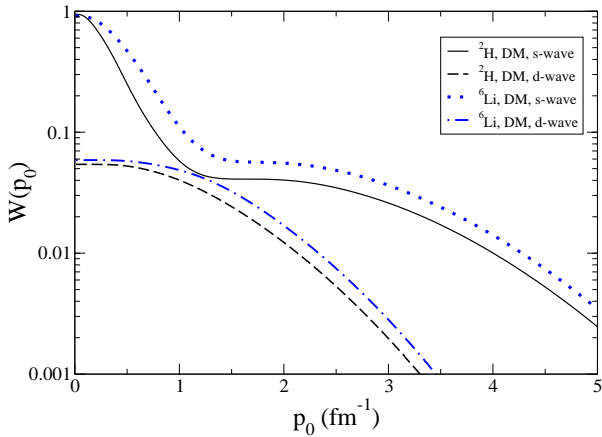}
\caption{Probability of the high-momentum $S$- and $D$-wave components with $(p\geqslant p_{0})$ in the nucleon momentum distributions for the $^{6}$Li nucleus (dotted and dot-dashed curves) and the deuteron (solid and dashed curves) obtained with the dibaryon model for the $NN$ and three-body forces.}
\end{center}
\label{w_Li-d}
\end{figure}
In Fig.~7 we present these probabilities for the $S$- and $D$-wave components of the wave functions of the $^{6}$Li nucleus and the deuteron found with the dibaryon-induced interaction. It is seen from the Figure that at $p_{0} \gtrsim 1.5\, {\rm fm}^{-1}$ (300 MeV$/c$)
the weight of the high-momentum $np$ components in $^6$Li is $\sim 40\%$ greater than that in deuteron both for $S$- and $D$-waves.

The probabilities of the high-momentum $NN$ components at $p_{0} = 1.5$ fm$^{-1}$ ($W_\lambda(1.5)$)
in the $^2$H, $^6$Li, and $^6$He nuclei are presented in Table~\ref{whm}.
The total weight $(w_{s}+w_{d})$ of the high-momentum components with $p>300 \,{\rm MeV}/c$ in the free deuteron for the dibaryon-induced model
is 6.5\% vs 4.3\% for the conventional (RSC) model.
\begin{table}[h!]
\caption{The probabilities $W_\lambda$ (in percent) of the high-momentum $NN$ components of the $^2$H, $^6$Li, and $^6$He wave functions at $p_{0} = 1.5$ fm$^{-1}$}
\begin{center}
\begin{tabular}{ccccccc}
\hline
nucleus &\multicolumn{2}{c}{ $^2$H}&\multicolumn{2}{c}{ $^6$Li}&\multicolumn{2}{c}{ $^6$He}\\ \hline
Model      & DM& RSC& DM & RSC &DM& RSC\\ \hline
$W_s$ &    4.10   &   0.96     &   5.84	   &	1.32	 &  3.86      &  0.54        \\ \hline
$W_d$ &    2.42   &   3.38     &   3.22	   &	4.07	 &	      & \\ \hline
$W_s+W_d$& 6.52   &   4.34     &   9.06	   &	5.39	 &	       & \\ \hline
\end{tabular}
\end{center}
\label{whm}
\end{table}

An especially interesting comparison can be done for the weights $W_\lambda$ of the $S$- and $D$-wave components in the $^6$Li, $^6$He, and $^2$H nuclei calculated within the conventional and dibaryon-induced force models (see Fig.~8 and Table~\ref{whm}). We see that $W^{RSC}_{d}> W^{RSC}_{s}$ while $W^{DM}_{d}
< W^{DM}_{s}$ at $p_0=300 \,{\rm MeV}/c$.
In other words, for the conventional force model the $D$-wave component gives the main contribution to the nucleon high-momentum distribution both in the deuteron and $^6$Li (this is a well-known fact \cite{Wiringa}), while for the dibaryon-induced model {\em the situation is opposite}: the $S$-wave component gives the main contribution to nucleon momentum distribution for both nuclei at $p_{0}\gtrsim 300 \,{\rm MeV}/c$.
\begin{figure}[h!]
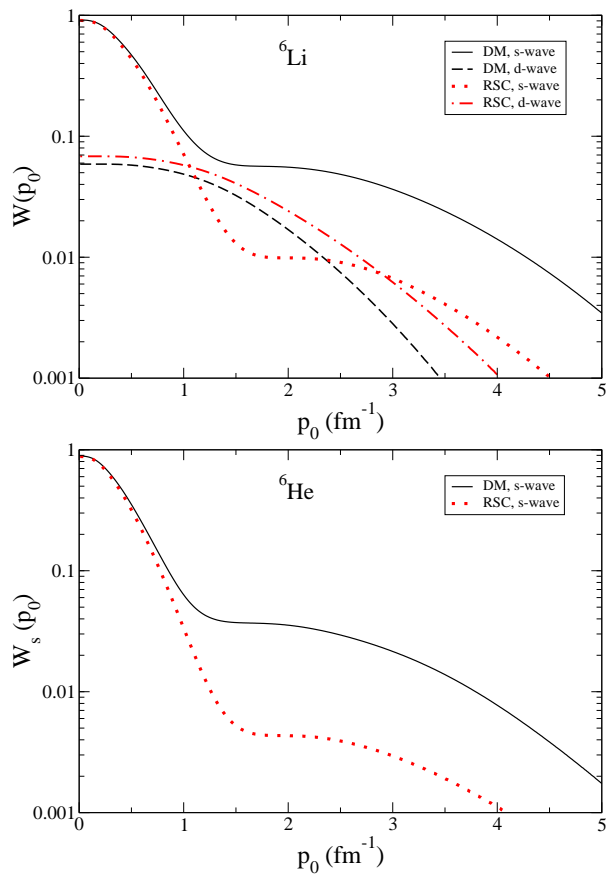

\begin{center}
\includegraphics[width=0.45\textwidth]{fig8a.eps}
\includegraphics[width=0.45\textwidth]{fig8b.eps}
\caption{Probability of the high-momentum components with $(p\geqslant p_{0})$ in the nucleon momentum distribution for $^{6}$Li and $^{6}$He calculated with the RSC potential and the dibaryon model.}
\end{center}
\label{w_Li6}
\end{figure}

The enhancement of the short-range ($S$-wave) neutron-neutron correlations in case of the dibaryon-induced interaction vs. the conventional (RSC) $nn$ interaction can be most clearly seen in the $^6$He nucleus (see Fig.~8, bottom panel, and
Table~\ref{whm}). Here the probability of the high-momentum $nn$ components at $p>300
\,{\rm MeV}/c$ for the dibaryon-induced model is seven times larger than that for the
traditional RSC model. Such a strong enhancement of the short-range correlations is
one of the most noticeable distinctions of our approach to $NN$ interaction from the
conventional ones.

\subsection {Substantiation of the cluster model}
Before passing to a general discussion of the results obtained it is worth considering some important arguments in favor of the validity of the three-cluster $\alpha-2N$ model for the description of the $A=6$ nuclei.

\begin{figure}[h!]
\center{\includegraphics[width=0.95\columnwidth]{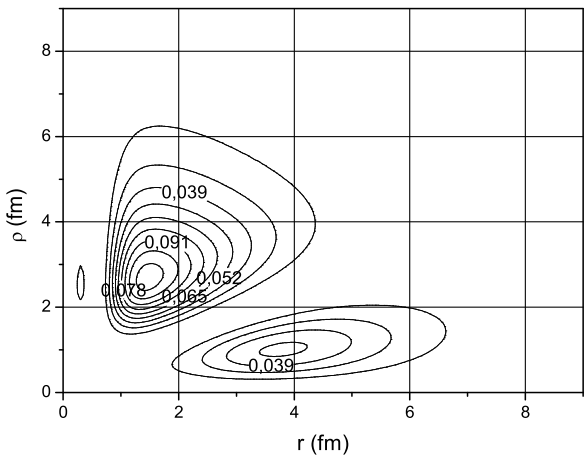} }
\caption{Isolines for the probability density $W(\rho ,r)$ for the $\gamma=(0001)$ component of the $^6$Li($1^+0$)-state wave function calculated within the $\alpha+2N$ cluster model with the dibaryon-induced $NN$ and three-body interactions. The coordinate $r$ means the internucleon distance, while $\rho$ is the distance between the $NN$ center of mass and the $\alpha$-cluster.}
\label{g_Li6}
\end{figure}

\begin{figure}[h]
\center{\includegraphics[width=0.95\columnwidth]{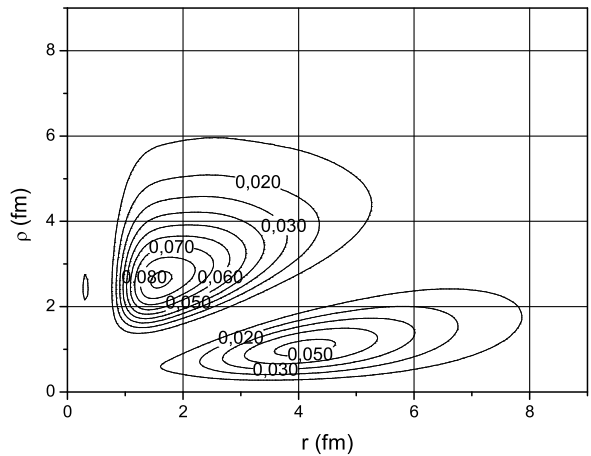} }
\caption{The same as in Fig.\ref{g_Li6} but for the $\gamma=(0000)$ component of the $^6$He($0^+1$) state.}
\label{g_He6}
\end{figure}

First of all, let us consider the geometrical forms and the matter distribution in the $^6$Li and $^6$He nuclei found in our calculations (see Figs. \ref{g_Li6} and \ref{g_He6}).
%\subsection{Geometric shapes of the ground states of $^{6}$Li and $^{6}$He nuclei}

Figs. \ref{g_Li6} and \ref{g_He6} present the isodensity
levels for the probability density
\begin{equation}
 W_{\gamma}(r,\rho ) = |\Phi _{\gamma}^J(r,\rho )|^2 r^2 \rho ^2 /P_\gamma,
 \label{density}
\end{equation}
calculated for the main $S$-wave components of the ground states of $^{6}$Li and $^{6}$He
within the dibaryon-induced model for $NN$ and three-body $\alpha NN$ forces.
In Eq.(\ref{density}) $P_\gamma $  is the weight of the $\gamma$-component, so the probability density is normalized to unity: $\int W(r,\rho ) drd\rho=1$.  Therefore, the density value $W_\gamma $ at maximum, marked by
numerals on the chart of isolines, allows us to estimate the sharpness
of the peaks.

 In Refs.~\cite{NP90,NP93,NP95} the geometric shapes of the $ A =6$  nuclei  were
predicted and investigated in detail for the case of the traditional $NN$ force.
It has been shown that there are three  basic geometric structures, each of which
can be described by the factorized wave function:  the cluster  $ \alpha-d$
configuration (``dumb-bell''), the ``cigar'' shape with an  $\alpha$-particle
on the straight line between two nucleons, and the ``helicopter'' configuration with the double
rotation. These predictions were completely confirmed both in the experiments
\cite{Bochkarev} and in a series of calculations \cite{Danilin} using the $K$-harmonic
method. In particular, it was found that the  $S$-components  $ (\lambda =0,
l=0) $  always contain two structures: cluster and cigar-like  with the peak
height ratio $ 1.7\div 2$.

 It can be seen from Figs. \ref{g_Li6} and \ref{g_He6} that replacing the
traditional $NN$ potential by the dibaryon-induced one does not
substantially change the geometrical shape of the nuclei. The Figures also clearly show that the three particles ($N,N,\alpha$) are localized
in the $^{6}$Li and $^{6}$He nuclei rather far from each other as compared with the own
dibaryon size ($r_D \sim 0.6 - 0.9$~fm), and the distance between the $2N$ center of mass and the
$\alpha$-particle in the cluster configuration is about 3~fm. This distance is much
larger than the range of the OSE three-body force ($r_{\rm OSE}\simeq 0.56$~fm for $m_\sigma \simeq 350$~MeV) and much larger than the
range of the TSE interaction.

Therefore the three-body force contribution is very sensitive to the $\sigma NN$
coupling constant. This leads to a highly nonlinear effect for the dependence of the
three-body binding energy on the dibaryon contribution. As a result of this non-linear
effect, the peak density in two basic configurations of $^6$Li is greater than that
of $^6$He (see Figs.~\ref{g_Li6} and \ref{g_He6}).

From the density profiles presented in Figs.~\ref{g_Li6} and \ref{g_He6}, one can
see that the average distances between three particles in the $\alpha-N-N$ system are much
larger than the ranges of the respective interactions $V_{NN}$ and $V_{\alpha N}$, so that,
the overlap between the outer and $\alpha$-particle nucleons is very small ($4$--$5$\%) and
the possible distortion of the $\alpha$-cluster in these nuclei should be minimal and
can be ignored. This conclusion can serve as an additional good confirmation for the
applicability of the three-cluster model for the $A=6$ nuclei used in this paper.

\section{Dibaryon admixture at different nuclear densities}

The weight of the dibaryon components in the nuclear wave functions is of great interest because many properties of nuclear matter can depend crucially upon the dibaryon admixture.
The most natural way to change the dibaryon admixture is to change the binding energy of the $NN$ pair in the given nucleus. It can be done easily by increasing the $g_{\sigma NN}$ coupling constant which controls the contribution of the three-body force due to OSE.

Within the framework of the cluster model considered in this paper, we studied
the dependence of the dibaryon admixture on the average distance
between the nucleons forming the dibaryon, i.e., on the density of the nuclear
medium. To do this, we carried out a series of calculations, increasing the
magnitude of the three-particle force between the pair of nucleons and the $\alpha$-core
which is determined by the coupling constant $g_{\sigma NN}$ of OSE.

\begin{figure}[h!]
\center{\includegraphics[width=0.95\columnwidth]{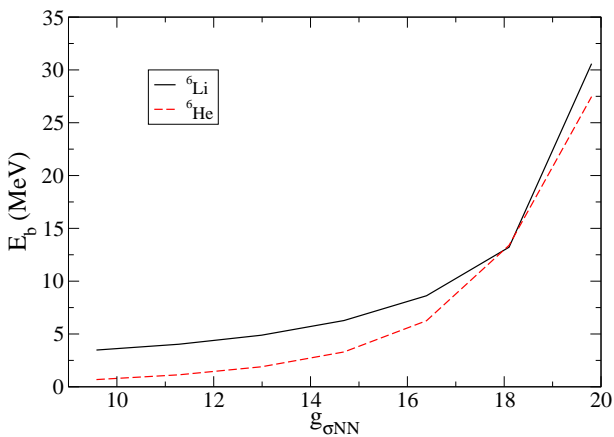} }
\caption{Dependence of the binding energy $E_b$ for $^{6}$Li and $^{6}$He on the coupling constant $g_{\sigma NN}$ which determines the strength of the OSE three-body force. }
\label{Eb}
\end{figure}

\begin{figure}[h!]
\center{\includegraphics[width=0.95\columnwidth]{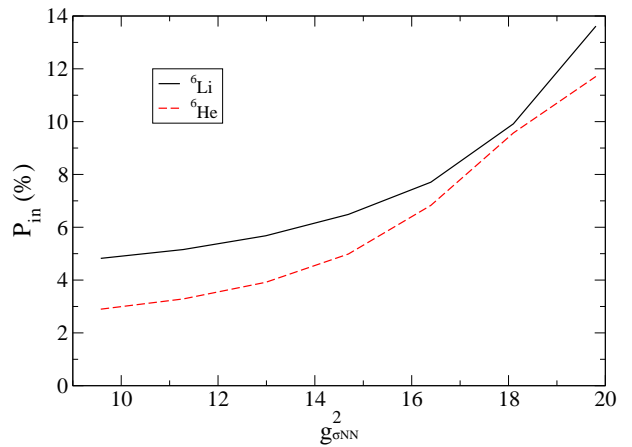} }
\caption{Dependence of the dibaryon admixture $P_{\rm in}$ in $^{6}$Li and $^{6}$He on the  coupling constant $g_{\sigma NN}$. }
\label{Pb}
\end{figure}

The results of this study for both $^{6}$Li and $^{6}$He nuclei are presented in
Figs.~\ref{Eb} and \ref{Pb}.

From Fig.~\ref{Eb} it is seen clearly that the behavior of $E_b$ as a function
of $g_{\sigma NN}$ is rather similar for the $^{6}$Li and $^{6}$He nuclei at the moderate
values of $g_{\sigma NN}$, but for high values of $g_{\sigma NN}$ the binding
energy for two neutrons in the field of the $\alpha$-core rises faster than for the
neutron-proton pair.

The similar trend is observed for the weight of the dibaryon component in the total
three-cluster wave functions for $^{6}$Li and $^{6}$He: the admixture $P_{\rm in}(nn)$ of the neutral
$nn$ isovector dibaryon in $^{6}$He is lower than the admixture $P_{\rm in}(np)$ of
the charged $np$ isoscalar dibaryon in $^{6}$Li at the ``physical'' value
of the constant $g_{\sigma NN}$, but rises faster than $P_{\rm in}(np)$ with increasing
$g_{\sigma NN}$. At large values of $g_{\sigma NN}$ these admixtures become
close to each other (see Fig.~\ref{Pb}).

As is seen from Figs. \ref{Eb} and \ref{Pb}, when the binding of the $NN$ pair inside the
nucleus becomes stronger, the average $NN$ distance gets smaller and thus the
admixture of the dibaryon component gets higher.  If to assume that the binding
energy of two nucleons in the middle-weight nuclei ($A\sim 50$) is ca. 16~MeV, then
this binding energy in our model corresponds to the value $g_{\sigma NN} \simeq 19$
at which the dibaryon admixture $P_{\rm in}$ is about 11\%. This value is in very
good agreement with our previous estimations for the dibaryon weight in the $A=3$
nuclei: $P_{\rm in}(3N) =10$--$12$\% (see Tab. 3). So, it means that only $1/10$ of all
$np$ pairs can form the dibaryons in a nucleus simultaneously. Thus we can neglect
the interaction between the different dibaryons inside the nucleus and consider them as
independent two-nucleon clusters.

\begin{figure}[h!]
\center{\includegraphics[width=0.95\columnwidth]{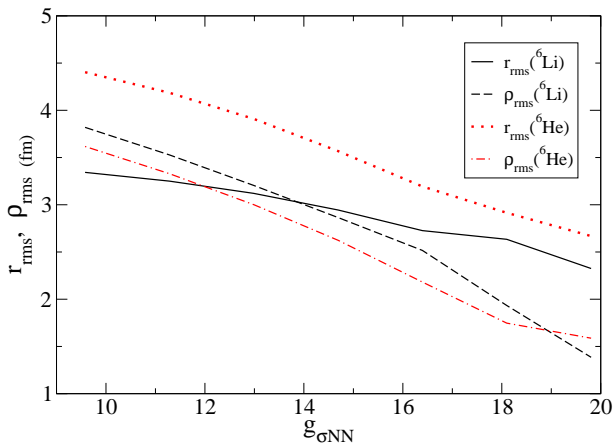} }
\caption{The rms value for the internucleon distance $\sqrt{\langle r^2 \rangle}$ and the rms value for the distance between the $\alpha$-particle and the $2N$ center of mass $\sqrt{\langle \rho^2 \rangle}$ as functions of the coupling constant $g_{\sigma NN}$. }
\label{rms_xy}
\end{figure}

Other interesting quantities to study is the rms internucleon distance, $r_{\rm
rms}$, and the rms distance between the $\alpha$-particle and the $2N$ center of mass, $\rho_{\rm
rms}$. The respective rms values for $^{6}$Li and $^{6}$He are displayed in
Fig.~\ref{rms_xy} as functions of $g_{\sigma NN}$.
From the Figure one can conclude
that while the rms internucleon distance $r_{\rm rms}$ in $^{6}$He is greater than in
$^{6}$Li, the opposite behavior is observed for the rms $\alpha-NN$ distance
$\rho_{\rm rms}$. Therefore, the rms radii of the matter distribution
$\sqrt{r^{2}_{\rm m}}$ in both nuclei turn out to be very close to each other (see
Tab.~\ref{properties}).

\section{Conclusion}

The present paper represents the first study in the literature of the properties
of the low-lying dibaryon resonances, namely the $np$ $^3S_1$ and $nn$ (or $pp$) $^1S_0$ resonances predicted by Dyson and
Xuong~\cite{Dyson} as early as 1964, in nuclear medium. We presented the results of two series of
calculations for the $A=6$ nuclei $^{6}$Li and $^{6}$He within the three-cluster $\alpha+2N$
model for two alternative types of the basic $NN$ interaction: the conventional (RSC) $NN$
potential and the dibaryon-induced model including the effective pair $NN$ potential
together with the new three-body force arising due to the interaction of the intermediate dibaryon
with the nucleons in the $\alpha$-core. We compared the results for the properties
of the $^{6}$Li and $^{6}$He nuclei obtained with the use of two above $NN$-force models with each other and with the experimental data.

Our main conclusion is as follows: in many aspects, the properties of the real $^{6}$Li
and $^{6}$He nuclei, e.g., the binding energies, the rms
radii of the charge and matter distributions, etc., are reproduced better with the dibaryon-induced model than with
the traditional RSC $NN$ interaction. We found also that the
neutron-neutron short-range correlations in $^{6}$He are enhanced in case of the
dibaryon-induced interaction as compared to the conventional force model. It may
give a key to a better understanding of $nn$ pairing in nuclei at all.

In the present work, we also investigated the properties of the $nn$ and $np$
dibaryons in the field of the $\alpha$-core and their dependence on the magnitude of the
three-particle force due to the interaction of the dibaryon with the $\alpha$-core.

There is another important aspect of the present study. A large series of
recent works worldwide have been devoted to the short-range correlations (SRC) of nucleons in
nuclei. It is now generally accepted that the SRC are based on fluctuations in the
density of nucleons in nuclei. The generation of di-, tri-, or multibaryons in
nuclei resulting from the dibaryon concept of nuclear force can be considered as a
source of such density fluctuations and, therefore, of the SRC.

We summarize with the hope that the present study will motivate researchers in the field over the world to study in detail other manifestations of the dibaryon degrees of freedom in nuclei and nuclear medium.

{\bf Acknowledgments.} The authors are grateful to Prof. Elena Zemlyanaya for her help in preparing the computer code used to obtain the  numerical results presented in the paper. The work has been partially supported by the Russian Foundation for Basic Research, grants Nos. 19-02-00011 and 19-02-00014.


\begin{thebibliography}{ab}
\bibitem{Yukawa} H. Yukawa, Proc. Phys. Math. Soc. Japan {\bf 17}, 48 (1935).
\bibitem{Breit} K.F. Lassila, M.H. Hull Jr., Y.M. Ruppel, F.A. McDonald, G. Breit, Phys. Rev. {\bf 126}, 881 (1962).
\bibitem{Signell} P.S. Signell, R.E. Marshak, Phys. Rev. {\bf 109}, 1229 (1958);
P.S. Signell, R Zinn, R.E. Marshak, Phys. Rev. Lett. {\bf 1}, 416 (1958).
\bibitem{Bonn_NN} R. Machleidt and I. Slaus, J. Phys. G {\bf 27}, R69 (2001).
\bibitem{5}  W. Gl\"ockle, H. Wita\l{}a, D. H\"uber, H. Kamada, and J. Golak, Phys. Rept. {\bf 274}, 107 (1996).
\bibitem{6} H. Wita\l{}a, J. Golak, R. Skribinski, W. Gl\"ockle, W.N. Polyzou, and H. Kamada, Phys. Rev. C {\bf 77}, 034004 (2008).
\bibitem{7} V.I. Kukulin, M.N. Platonova, Phys. At. Nucl. {\bf 76}, 1465 (2013).
\bibitem{8} K. Sagara, Few-Body Syst. {\bf 48}, 59 (2010).
\bibitem{9} K. Sekiguchi {\em et al.}, Phys. Rev. C {\bf 79}, 054008 (2009).
\bibitem{10} R. Machleidt, Adv. Nucl. Phys. {\bf 19}, 189 (1989).
\bibitem{Weinberg} S. Weinberg, Phys. Lett. B {\bf 251}, 288 (1990); Nucl. Phys. B {\bf 363}, 3 (1991).
\bibitem{Machl} R. Machleidt and D. R. Entem, Phys. Rept. {\bf 503}, 1 (2011).
\bibitem{Eppelbaum} E. Epelbaum and J. Gegelia, Eur. Phys. J. A {\bf 41}, 341 (2009).
\bibitem{Ordones} C. Ord\'o\~nes, L. Ray, and U. van Kolck, Phys. Rev. Lett. {\bf 72}, 1982 (1994); Phys. Rev. C {\bf 53}, 2086 (1996).
\bibitem{Hammer} H.-W. Hammer, S. K\"onig, and U. van Kolck, Rev. Mod. Phys. {\bf 92} 025004 (2020).
\bibitem{Margaryan} A. Margaryan, P.R. Springer, and J. Vanasse, Phys. Rev. C {\bf 93}, 054001 (2016).
\bibitem{Leksin} G. A. Leksin, Phys. At. Nucl. {\bf 65}, 1985 (2002);
in : {\em Elementary particles. III ITEF Physics School} (Atomizdat, Moscow, 1975), iss. 2, p. 5.
\bibitem{16} A.V. Efremov, Fiz. Elem. Chast. At. Yadra {\bf 13}, 613 (1982).
\bibitem{17} V.S. Stavinsky, Fiz. Elem. Chast. At. Yadra {\bf 10}, 949 (1979); JINR Rapid Communications, {\bf 18}, 865 (1986).
\bibitem{18} L. Kondratyuk, M. Shmatikov, Preprint ITEP-33 (1984).
\bibitem{19} U. Mosel, Subthreshold Particle Production in Heavy-Ion Collisions, Lecture at the V J.A. Swieca Summer School: Nuclear Physics, Caxambu, Brasil, 19--28 February 1991.
\bibitem{20} Z.-S. Wang, Y.-K. Ho, Z.-Y. Pan, J. Phys. G {\bf 20}, 1901 (1994).
\bibitem{21} V.S. Koptev {\em et al.}, JETP {\bf 94}, 1 (1988).
\bibitem{Hen} O. Hen, D.W. Higinbotham, G.A. Miller, E. Piasetzky, and L.B. Weinstein, Int. J. Mod. Phys. E {\bf 22}, 133017 (2013).
\bibitem{23} L. Weinstein, ``Where are the quarks?'', Seven Years of Nuclear Physics at CEBAF, talk at the Annual CEBAF users group meeting, 2003; L. Weinstein {\em et al.}, Phys. Rev. Lett. {\bf 106}, 052301 (2011).
\bibitem{24} M. Sargsyan, ``QCD and Nuclear Physics'', talk at the Annual CEBAF users group meeting, 2003;  O. Hen, G. A. Miller, E. Piasetzky, and L. B. Weinstein, Rev. Mod. Phys. {\bf 89}, 045002 (2017).
\bibitem{PIYAF} V.I. Kukulin, in: {\em Proceedings of XXXIII Winter School PIYaF}, Gatchina, 1999, p. 207.
\bibitem{YaF2001} V.I. Kukulin, I.T. Obukhovsky, V.N. Pomerantsev, and
A. Faessler, Phys. At. Nucl. {\bf 64}, 1667 (2001).
\bibitem{JPhys2001}  V.I. Kukulin, I.T. Obukhovsky, V.N. Pomerantsev, and
A. Faessler, J. Phys. G {\bf 27}, 1851 (2001).
\bibitem{KuInt} V.I. Kukulin, I.T. Obukhovsky, V.N. Pomerantsev, and
A. Faessler, Int.~J. Mod.~Phys.~E {\bf 11}, 1 (2002).
\bibitem{Yaf2019} V.I. Kukulin, V.N. Pomerantsev, O.A. Rubtsova, and M.N. Platonova, Phys. At. Nucl. {\bf 82}, 934 (2019).
\bibitem{PLB} V.I. Kukulin, O.A. Rubtsova, M.N. Platonova, V.N. Pomerantsev, H. Clement, Phys. Lett. B {\bf 801}, 135146 (2020).
\bibitem{EPJA} V.I. Kukulin, O.A. Rubtsova, M.N. Platonova, V.N. Pomerantsev, H. Clement, T. Skorodko, Eur. Phys. J. A {\bf 56}, 229 (2020).
\bibitem{sys3n} V.I. Kukulin, V.N. Pomerantsev, M. Kaskulov, and
A. Faessler, J.~Phys.~G {\bf 30}, 287 (2004);
V.I.~Kukulin, V.N.~Pomerantsev, and A.~Faessler, J.~Phys.~G {\bf 30}, 309 (2004).
\bibitem{YAF3N} V.N. Pomerantsev, V.I. Kukulin, V.T. Voronchev, and A. Faessler, Phys. At. Nucl. {\bf 68}, 1453 (2005).
\bibitem{voronchev82} V.T. Voronchev, V.M. Krasnopolsky, and V.I. Kukulin, J. Phys. G {\bf 8}, 649; 667 (1982).
\bibitem{NP84} V.I. Kukulin, V.M. Krasnopolsky, V.T. Voronchev, and
P.B. Sazonov, Nucl. Phys. {\bf A417}, 128 (1984).
\bibitem{NP86} V.I. Kukulin, V.M.  Krasnopolsky, V.T. Voronchev, and P.B. Sazonov, Nucl. Phys. {\bf A453}, 365 (1986).
\bibitem{NP90} V.I. Kukulin, V.T. Voronchev, T.D. Kaipov, and R.A. Eramzhyan, Nucl. Phys.{\bf A517}, 221 (1990).
\bibitem{NP93} G.G.  Ryzhikh, R.A. Eramzhyan, V.I. Kukulin, and Yu.M.
Tchuvilsky, Nucl. Phys. {\bf A563}, 247 (1993).
\bibitem{NP95} V.I. Kukulin, V.N. Pomerantsev, Kh.D. Razikov, V.T. Voronchev, G.G. Ryzhikh, Nucl. Phys. {\bf A586}, 151 (1995).
\bibitem{Rel} H. Kamada, W. Gloeckle, J. Golak, and Ch. Elster, Phys. Rev. C
{\bf 66}, 044001 (2002).
\bibitem{Danilin}  B.V. Danilin {\em et al.}, Yad. Fizika {\bf 49}, 351; 360 (1989). (in Russian)
\bibitem{40} B.V. Danilin {\em et al.}, Yad. Fizika  {\bf 48}, 1208 (1988). (in Russian)
\bibitem{41} B.V. Danilin, M.V. Zhukov, A.A. Korsheninnikov and
L.V. Chulkov, Yad. Fizika  {\bf 53}, 71 (1991) (in Russian);\\
B.V. Danilin, M.V. Zhukov, S.N. Ershov, F.A. Gareev, R.S.
Kurmanov, J.S. Vaagen and J.M. Bang, Phys. Rev. C {\bf 43}, 2835 (1991).
\bibitem{Lawson} R.D. Lawson, Nucl. Phys. {\bf A148}, 401 (1970).
\bibitem{Charl} C. Ji, Ch. Elster, D. R. Phillips, Phys. Rev. C {\bf 90}, 044004 (2014)

\bibitem{PLB83} V.M. Krasnopolsky, V.I. Kukulin, P.B. Sazonov and V.T. Voronchev, Phys. Lett. B {\bf 121}, 96 (1983).
\bibitem{Varga} Y. Suzuki, K. Varga, {\em Stochastic variational approach to quantum mechanical few-body problems}, Springer, 1988.
\bibitem{SAID} R.A. Arndt, W.J. Briscoe, I.I. Strakovsky, R.L. Workman, Phys. Rev. C {\bf 76}, 025209 (2007). All SAID PWA solutions can be accessed via the official SAID website: http://gwdac.phys.gwu.edu
\bibitem{FBS2019} V.N. Pomerantsev, V.I. Kukulin, O.A. Rubtsova, Few-Body Syst. {\bf 60}, 48 (2019).
\bibitem{Pieper2001} S.C. Pieper, V.R. Pandharipande, R.B. Wiringa, and J.~Carlson, Phys. Rev. C {\bf 64}, 014001 (2001).
\bibitem{Nogga2003} A. Nogga {\em et al.}, Phys. Rev. C {\bf 67}, 034004 (2003).
\bibitem{Tanihata92} I. Tanihata, D. Hirata {\em et al.}, Phys. Lett. B {\bf 289}, 261 (1992).
\bibitem{Horiuchi} W. Horiuchi and Y. Suzuki, Phys. Rev. C {\bf 76}, 024311 (2007).

\bibitem{J-Lab1} L. B. Weinstein and R. Niyazov, Eur. Phys. J. A {\bf 17}, 429 (2003).
\bibitem{J-Lab2} R. Subedi, R. Shneor, P. Monaghan, B. D. Anderson, K. Aniol, J. Annand,
J. Arrington, H. Benaoum, F. Benmokhtar, W. Boeglin, J.-P. Chen, Seonho Choi, E. Cisbani, B. Craver, S. Frullani, F. Garibaldi, Science {\bf 320}, 1476 (2008).
\bibitem{J-Lab3} N. Fomin, J. Arrington, R. Asaturyan, F. Benmokhtar, W. Boeglin, P. Bosted, A. Bruell, M. H. S. Bukhari, M. E. Christy, E. Chudakov, B. Clasie, S. H. Connell, M. M. Dalton,
A. Daniel, D. B. Day, D. Dutta, Phys. Rev. Lett. {\bf 108}, 092502 (2012).
\bibitem{J-Lab4} I. Korover {\em et al.} (Jefferson Lab Hall A Collab.), Phys. Rev. Lett. {\bf 113}, 022501 (2014).

\bibitem{Blomq} K.I. Blomqvist {\em et al.}, Phys. Lett. B {\bf 424}, 33 (1998).
\bibitem{Wiringa} R.B. Wiringa, R. Schiavilla, Steven C. Pieper, and J. Carlson,
Phys. Rev. C {\bf 89}, 024305 (2014).
\bibitem{Bochkarev} O.V. Bochkarev {\em et al.}, Pisma v Zh. Eks. Teor. Fiz. {\bf 40}, (1984) (in Russian);
O.V. Bochkarev {\em et al.}, Yad. Fizika {\bf 46}, 12 (1987). (in Russian)
\bibitem{Dyson} F.J. Dyson and N.-H. Xuong, Phys. Rev. Lett. {\bf 13}, 815 (1964).

\end{thebibliography}
\end{document}